\documentclass{article}
\usepackage{amssymb,color,graphicx}
\usepackage{placeins}
\usepackage{amssymb}
\usepackage{wrapfig,framed,cancel}
\usepackage[colorlinks=true, pdfstartview=FitV, linkcolor=black, citecolor=black, urlcolor=blue]{hyperref}
\definecolor{shadecolor}{rgb}{0.93, 0.93, 0.86}
\newtheorem{theorem}{Theorem}[section]
\newtheorem{prop}{Proposition}[section]

\newtheorem{defn}{Definition}[section]

\newtheorem{remark}{Remark}[section]

\def\s{\sigma}
\def\br{\begin{remark}\rm}
\def\er{\end{remark}}

\def \bd{\begin{defn}}
\def\ed{\end{defn}}
\def\l{\lambda}
\def\C{{\mathbb C}}

\def\bR{{\mathbb R}}
\def\AR{\mathcal A}
\def\eqref#1{ (\ref{#1})}
\def\&{&\hspace{-20pt}}
\def\R{{\mathbb R}}
\def\d{{\rm d}}

\def\1{\mathbf 1}
\def\a{{\alpha}}

\def\wh{\widehat}

\def\wt{\widetilde}

\def\bea{\begin{eqnarray}}
\def\eea{\end{eqnarray}}
\def\Id{\mathrm{Id}}
\def\Ai{\mathrm{Ai}}

\def\g{{\gamma}}

\def\0{{\bf 0}}

\def\Ai{\mathrm{Ai}}

\def\QED{ {\bf Q.E.D}\par \vskip 4pt}

\newcommand{\be}{\begin{eqnarray}}
\newcommand{\ee}{\end{eqnarray}}
\newcommand{\bes}{\begin{eqnarray*}}
\newcommand{\ees}{\end{eqnarray*}}

\definecolor{light-blue}{rgb}{0.8,0.85,1}
\definecolor{blue}{rgb}{0,0,1}
\definecolor{red}{rgb}{1,0,0}

\def\le{\left}
\def\ri{\right}
\def\ba{\begin{eqnarray}}
\def\eeq{\end{eqnarray}}

\makeatletter
\@addtoreset{equation}{section}
\makeatother

\begin{document}

\baselineskip 15pt plus 1pt minus 2pt
\begin{flushright}
\end{flushright}
\vspace{0.2cm}
\begin{center}
\begin{Large}
\textbf{The gap probabilities of the tacnode, Pearcey and Airy point processes, their mutual relationship and evaluation}
\end{Large}

\bigskip
M. Bertola$^{\dagger\ddagger}$\footnote{bertola@crm.umontreal.ca},  
M. Cafasso$^{\diamondsuit}$ \footnote{cafasso@math.univ-angers.fr}
\\
\bigskip
\begin{small}
$^{\dagger}$ {\em Centre de recherches math\'ematiques,
Universit\'e de Montr\'eal\\ C.~P.~6128, succ. centre ville, Montr\'eal,
Qu\'ebec, Canada H3C 3J7} \\
\smallskip
$^{\ddagger}$ {\em  Department of Mathematics and
Statistics, Concordia University\\ 1455 de Maisonneuve W., Montr\'eal, Qu\'ebec,
Canada H3G 1M8} \\
\smallskip
$^{\diamondsuit}$ {\em LUNAM Universit\'e, LAREMA, Universit\'e d'Angers\\ 2 Boulevard Lavoisier, 49045 Angers, France.}\\
\end{small}
\end{center}
\bigskip
\begin{center}{\bf Abstract}\\
\end{center}
We express the gap probabilities of the tacnode process as the  ratio of two Fredholm determinants; the denominator is the standard Tracy-Widom distribution, while the numerator is the Fredholm determinant of a very explicit kernel constructed with  Airy functions and exponentials.  The formula allows us  to apply the theory of numerical evaluation of Fredholm determinants and thus produce numerical results for the gap probabilities. In particular we investigate numerically how, in different regimes, the Pearcey process degenerates to the Airy one, and the tacnode degenerates to the Pearcey and Airy ones. 

\section{Introduction}

The study of ``\emph{infinite--dimensional diffusions}" arising from the scaling limit of determinantal point processes attracted much attention in the last years, see for instance  \cite{OxRMBook}, chapters 6,10,11,38 and references therein. One of the most popular model has been introduced by Dyson in \cite{Dy:BrownianMotions}, where a dynamical version of the probability distributions arising in random matrix theory is defined. The idea is to study $N$ Brownian particles moving on the real line, conditioned not to intersect. When the starting and ending points are fixed and $N$ goes to infinity, the particles sweep out a certain region in space-time, whose shape depends on the number and relative position of the starting and ending points. Interesting infinite dimensional diffusions arise when studying the behavior of the particles near the boundary of this region. Near the points where this boundary is a smooth curve, the  fluctuations of the particles are described by the well known Airy process \cite{PrSp,JoDetProc,Johansson3}. Near a cusp singularity we are lead to the study of the Pearcey process \cite{TracyWidomPearcey,OkounkovReshetikhin}. The most recently studied case is the one of the so--called tacnode singularity, appearing when the boundary looks (locally) as two circles touching at one point. 
The first expression for the kernel of the tacnode process has been written in \cite{AFvMTacnode}, as a scaling limit of a model of random walks, continuous in time. Few months later, in \cite{DelKuiZhaTacnode}, the authors found a different expression (for the one--time case) in terms of a $4\times 4$ Riemann--Hilbert problem, this time starting from non--intersecting Brownian motions. Another formula for the multi--time case, again different from the previous ones, has been found by Johansson in \cite{JohanssonTacnode}. There, for the first time, the resolvent of the Airy kernel appeared in the formula of the tacnode kernel (this is quite an important point for our present work, as it will be apparent later). In \cite{AJvMTacnode}, finally, the authors analyzed the same process as arising from random tilings instead of Dyson Brownian motions: in this paper it has been proven that all the different formulations above are indeed equivalent, thus performing a significative step in the direction of universality for the tacnode process. A similar result has been obtained by Delvaux in \cite{DelvauxEqualitiesTacnodes}. A more general formulation of this process in the ``asymmetric'' case has been studied in \cite{FerrariVetoTacnode}.

While the kernels describing the Airy and Pearcey processes can be expressed as simple double integrals of exponential functions (both in the one and multi--time case), studying the tacnode process presents an additional complication, since its kernel is highly transcendental. Indeed, it is written in terms of the resolvent of the Airy operator. In particular, given the known representations of the tacnode kernel,  the study of the associated (joint) gap probability (i.e. the Fredholm determinant associated to the kernel) appears as quite complicated.

The purpose of this short investigation is to show how to express the gap probability of the tacnode process in terms of Fredholm determinants associated to {\em explicit} kernels, no more complicated than the standard Airy one. Explicitly the key theoretical result is Theorem \ref{main} that expresses the gap probability of the tacnode process as a ratio of determinants (notation to be defined therein)
\be
 {\rm Prob} \le\{ \mathcal T_\s(\tau_i) \not\in E^{(i)},\ \ i=1,\dots, r\ri\} = 
\frac {\det \le[ \Id - \mathbb H_E\ri]       } {  \det \le[\Id_{[\wt\s,\infty)} - \pi_{\wt \s} K_{\Ai} \pi_{\wt \s} \ri]}
\label{13}
\ee
where $\mathcal T_\s(\tau)$ denotes the tacnode point field at time $\tau$, the operator $\mathbb H$ acts on $$L^2\le(\R_+\sqcup \R_0\bigsqcup_{j=1}^r \R_{\tau_j} \ri)\simeq
L^2(\R_+)\oplus L^2(\R_0) \bigoplus_{j=1}^r L^2(\R_{\tau_j})$$ and the subscript $_E$ means that it is restricted to the following subspace 
$$ L^2(\R_+)\oplus L^2([\wt \s,\infty)) \oplus L^2\le(E^{(1)}\sqcup \dots \sqcup E^{(r)}\ri).$$
In the formula above, the subscripts on the various copies of $\R$ are simply to distinguish them from each other and remind that some are associated to the times $\tau_1,\dots, \tau_r$ and  each $E^{(j)},\ j=1,\dots, r$, is a finite union of bounded intervals.
The kernel $\mathbb H$ is described explicitly in Theorem \ref{main}, but here it suffices to say that it involves at most the standard Airy functions. In the one--time case it is a $3 \times 3$ kernel acting on $L^2(\bR_+)\oplus L^2(\bR)\oplus L^2(\bR)$ and reading as follows:
$$
\mathbb H(x,y) =  \left[
\begin{array} {c|c|c}
  0
& 
 -\Ai(x+y) 
&
\!\!  \Ai^{(-\tau)} (x\sqrt[3]{2}+\s-y  ) \!\!\!
 \\[10pt]
 \hline &&\\ 
 - \Ai  (x+y ) 
&
 0
 &
   \Ai^{(-\tau)} (x\sqrt[3]{2} + y-\s ) 
 \\[10pt]
 \hline &&\\
 \Ai^{(\tau)} (\s-x+y \sqrt[3]{2}  )\!\!
 &
\!  \Ai^{(\tau)} (x-\s+y\sqrt[3]{2} ) \!\!
  &
   0
\end{array}\right]
$$
where $\Ai^{(\tau)}(x) = 2^\frac 1  6{\rm e}^{ \tau x+\frac 2 3\tau ^3 }\Ai (x+\tau^2)$ and $\tau,\sigma$ are parameters of the tacnode process, respectively the time and the ``pressure'' (or overlap, see also below for details).
As an application of our result, we compute numerically, following the method of \cite{Bornemann}, the gap probability associated to the tacnode process, and its relationship with the Airy and Pearcey processes (a more detailed list of our results is given below). Theorem \ref{main} can also be used as a starting point to identify the tacnode gap probability with the tau function of a given Riemann-Hilbert problem, in the spirit of \cite{BertolaCafasso1,BertolaCafasso3}. A work in this direction is currently in preparation \cite{BCG1}. 

Before entering into the details of our results, let us spend few words about Fredholm determinants and their numerical evaluations.
Let $(X,\d\mu)$ be a (sigma-finite) measure space. Given an integral operator $\mathcal J$  on $L^2(X,\d \mu)$ (belonging to some trace ideal \cite{Si}) the computation of the (regularized) Fredholm-Carleman determinant is an essentially transcendental problem.
If the kernel $\mathbb J$ of the integral operator $\mathcal J$ is known and sufficiently regular, then the numerical evaluation of the determinant can be approached by a suitable approximation scheme which involves projecting onto a suitable finite dimensional subspace. In this way the Fredholm determinant can be approximated by a finite determinant.

%

The numerical evaluation of Fredholm determinants of operators on $\R$ (or disjoint unions of several copies thereof) has been recently shown to pose little obstacle \cite{Bornemann}; the gist of the beautiful idea is basically the simplest approach of discretization, paired with skillful use of numerical estimates associated with quadrature rules.

The effectiveness of the method is vastly improved if  the kernel is {\em explicit} in terms of functions that are already implemented or easily so.   The paramount example is  the Airy kernel and the corresponding Tracy-Widom \cite{TracyWidomLevel} distribution
\bea
K_{\Ai}(x,y):= \frac {\Ai(x) \Ai'(y)-\Ai'(x)\Ai(y)}{x-y}
\ \ F_2(s):= \det \le[\Id_{[s,\infty)} - \pi_s K_{\Ai} \pi_s\ri]\ ,\\ \ \ \pi_s:L^2(\R)\to L^2([s,\infty).\label{TWd}
\eea
As we said before, the kernel of the tacnode process is, however, highly transcendental; it is precisely with the results contained in Theorem \ref{main} that we make it directly amenable to the methods in \cite{Bornemann}.

The following sections are organized as follows:

\begin{itemize}
	\item In the second section, as a sort of warm-up, we show how to use the theoretical results in \cite{BertolaCafasso1} to compute (numerically) the gap probability of the Pearcey process. As an application, we give numerical confirmations of the degeneration of the Pearcey gap probability to a couple of Tracy--Widom distribution (see \cite{BertolaCafasso1} and, for a similar result, \cite{ACvM2}).
	\item In the third section, we start with our main theoretical result, i.e. the expression of the tacnode gap probability in terms of the ratio of two Fredholm determinants with simpler kernels. This allows us to show numerically, in the two subsections, how the tacnode gap probability degenerate, in different regimes, to the Pearcey and the Airy one. The degeneration tacnode-to-Pearcey has been already proven in \cite{GeudensZhangTacPearcey} and another proof will be given in \cite{BCG1}, together with the  tacnode-to-Airy one, which has not been proven yet (to the best of our knowledge).
	\item In the appendix we prove that, given a determinantal point process and a given subset of the state space, the point process conditioned not to have particles on that given subset is also determinantal, and we explicitly write its kernel (see also \cite{TaoWigner} for the case of point processes with a finite number of particles, and \cite{Lyons} for the discrete case). In the third section we will relate (formally) this result to our representation of the tacnode gap probability as a ratio of two Fredholm determinants.

\end{itemize}

\section{Numerical evaluation of the Pearcey gap probabilities}
Suppose that $I= [a_1,a_2]\cup \dots \cup[a_{2N-1}, a_{2N}]$ and denote with $\mathcal{P}$ the Pearcey process \cite{TracyWidomPearcey,OkounkovReshetikhin}; 
in \cite{BertolaCafasso1} it was proved that
\be
\mathbb P\le\{ \mathcal P(\tau) \not\in I \ri\} = \det \le[\1- \wt K_{\vec a}\ri]
\label{BC1}
\ee
where the Fredholm determinant on the right hand side is on the space $L^2(\g_L\cup i\R \cup \g_R)$ and the operator $\wt K$ has kernel $\wt K(\l,\mu)$ given by  (here $\chi_{_X}$ denotes the indicator function of a subset $X\subset \C$)
\bea\label{PearceyKernel}
\wt K_{\vec a}(\lambda,\mu) = \frac 1{2i\pi} \frac { {\rm e}^{\frac {\Theta_0(\lambda)-\Theta_0(\mu)}2} \chi_{_{i\R }}(\mu)\chi_{_{\gamma_R\cup \gamma_L}}(\lambda)  
- \sum_{j=1}^{2N}  (-)^j {\rm e}^{\frac {\Theta_{0}(\mu) -\Theta_0(\lambda)}2  + a_j \lambda-a_j\mu } \chi_{_{\gamma_R\cup \gamma_L}}(\mu)\chi_{_{i\R}}(\lambda)}{\lambda - \mu}\\
\Theta_0(\l):= \frac {\l^4}4 - \tau \frac {\l^2}2
\eea
The contours $\gamma_{R}$ is a contour in the right  halfplane that comes from infinity along $\arg \l= \frac \pi 4$ and returns to infinity along $\arg \l = -\frac \pi 4$ while $\gamma_L = -\gamma_R$. Our aim is to use the equivalent representation of the Percey gap probability given in \eqref{PearceyKernel} for numerical computations. 
For the case of multi-time Pearcey gap probabilities we refer to \cite{BertolaCafasso3}\footnote{We remark that in the cited reference the formul\ae\ (3.11, 3.12) should have an additional overall denominator $2i\pi$ and in the last line of formula (3.12) the alternating sign $(-1)^\ell $ should be $(-1)^{\ell +1}$. Similarly, the formul\ae\ (2.12, 2.13, 2.14) should all have an additional overall factor $\frac 1{2i\pi}$. }.
In order to apply Bornemann's procedure \cite{Bornemann} we prefer to have kernels acting on the interval $[0,1]$, hence we have chosen two hyperbolas parametrized  by 
\bea
\phi_R(s) = \frac 12 \le[\frac {s}{1-s} + \frac {1-s}s \ri] -  \frac i2 \le[\frac {s}{1-s} - \frac {1-s}s \ri]:(0,1)\to \gamma_R\\
\phi_L(s) = -\frac 12 \le[\frac {s}{1-s} + \frac {1-s}s \ri] +  \frac i2 \le[\frac {s}{1-s} - \frac {1-s}s \ri]:(0,1)\to \gamma_L\\
\phi_0(s) = \tan \le(\frac{2\pi s-\pi}2\ri):(0,1)\to i\R
\eea
and, pulling back the kernel $\wt K$ with the aid of the maps $\phi_{L,R,0}$, we obtained a kernel on $(0,1)^3$ (it can be extended to the closure by defining it as zero). Then the numerical evaluation of the resulting kernel can be carried out as in \cite{Bornemann}. We only point out that the Pearcey kernel involves special functions  (Pearcey integrals) that are {\em not} implemented in any library that we could find. Thus it seems that the equivalent representation \eqref{BC1} is convenient not only from the theoretical point of view to analyze the transition to the Airy process as in \cite{BertolaCafasso1}, but also from the point of view of effective numerical implementation.
\subsection{From the Pearcey to the Airy process}
Let us consider, for instance, the simplest gap probabilities for the Pearcey and Airy processes:
\bea
F_{\mathrm{P}}([a_p,b_p]; \tau) = \mathbb P\le\{\mathcal P(\tau)\not \in [a _p, b_p]\ri\}\ ,\qquad
F_{2}([\s,\infty)) = \mathbb P\le\{\mathcal A\not \in [\s,\infty) \ri\}
\eea

In \cite{BertolaCafasso1} it was shown by using the Deift--Zhou nonlinear steepest descent method that, in particular
\be
F_{\mathrm P}\le(\le[ -2\le(\frac \tau 3\ri)^\frac 3 2 + (3\tau)^\frac 1  6 \rho, 2\le(\frac \tau 3\ri)^\frac 3 2 - (3\tau)^\frac 1  6 \s,   \ri]; \tau \ri) \mathop{\longrightarrow}_{\tau \to\infty} F_{2}([\s,\infty)) F_{2}([\rho,\infty))
\label{PtoAi}
\ee
(the statement of \cite{BertolaCafasso1} was for an arbitrary finite union of bounded intervals, but here we restrict to the simplest nontrivial example). As an indirect confirmation that the numerical approach is sound, we verify this limit numerically plotting the graph of the ``\emph{relative difference}'' $ 1 - \frac {F_P([a_P, b_P];\tau)}{F_2(\s)F_2(\rho)}$ with the parameters as in \eqref{PtoAi} and $(\rho,\s)\in [-3,1]\times [-3,1]$. The two graphs in Figure 1 correspond to the case $\tau = 3$ and $\tau = 5.314$. In the second case, already for a relatively small value of $\tau$, the relative difference is below $15\%$ (in the range of space variables we considered).

\begin{figure}
\includegraphics[width=0.5\textwidth]{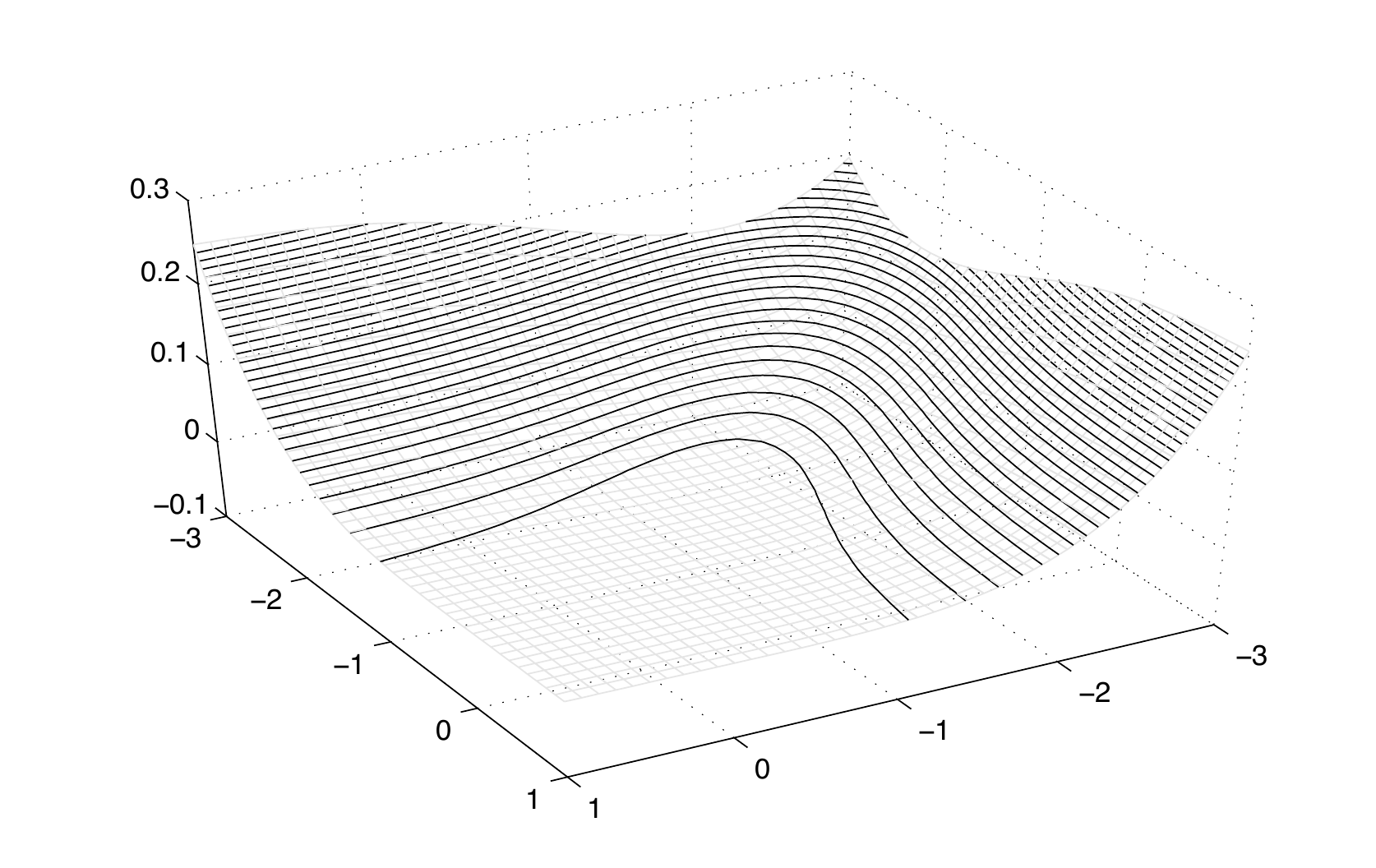}
\includegraphics[width=0.5\textwidth]{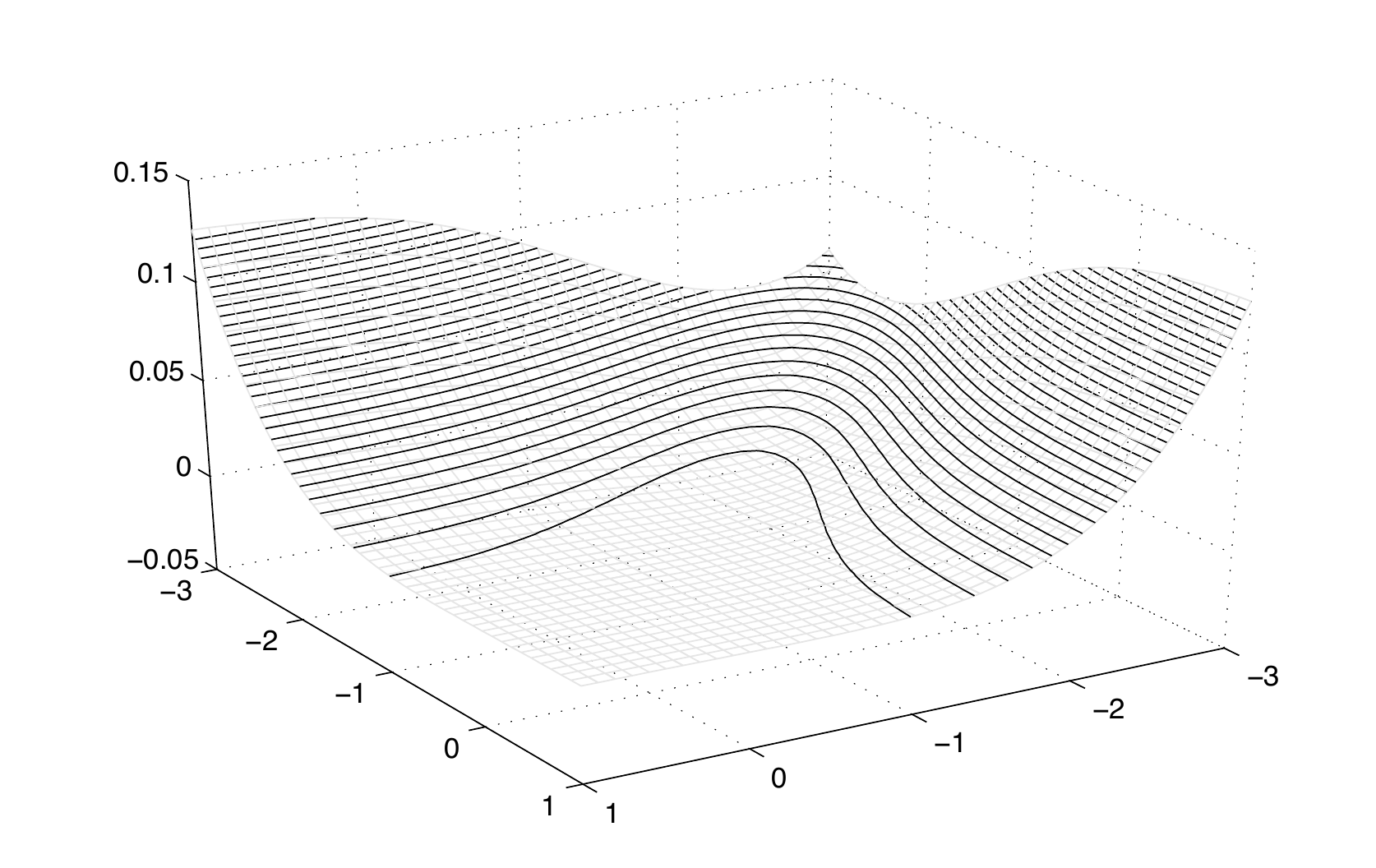}
\caption{\footnotesize The relative difference $ 1 - \frac {F_{\mathrm P}([a_P, b_P];\tau)}{F_2(\s)F_2(\rho)}$ with the parameters as in \eqref{PtoAi}, $(\rho,\s)\in [-3,1]\times [-3,1]$ and $\tau = 3$ (left) or $\tau = 5.314$ (right).}
\end{figure}
\FloatBarrier

\section{Gap probabilities of the tacnode process as ratio of determinants}

We start recalling one of the three equivalent formulations of the tacnode kernel obtained in \cite{AJvMTacnode}:

\bd[\cite{AJvMTacnode} formula (19)]
\label{defKtac}
The kernel of the  {\bf extended tacnode process with overlap $\sigma$} is defined as
 \bea
  {\mathbb K}^{\rm tac} (\tau_1, \xi_1;\tau_2,\xi_2 )  
= K_{\Ai}^{(\tau_1,-\tau_2)}(\s -\xi_1,\s-\xi_2)-\Id_{\tau_1>\tau_2}
p(\tau_1-\tau_2;\xi_1,\xi_2)+
\nonumber\\+
 \int_{\tilde\s}^{\infty}\left((\1- K_{\Ai}
 )_{\tilde\s}^{-1}
 \AR_{\xi_1-\s}^{ \tau_1}\right)(u)
\AR_{\xi_2-\s}^{ -\tau_2 }(u)du.
\label{E127}
\eea
where $\wt \s = 2^\frac 23 \s$ and  $K_{\Ai}$ denotes the usual Airy kernel and\footnote{For our convenience we slightly modified the definition of $\Ai^{(\tau)}$ multiplying it by the coefficient $2^{\frac{1}6}$. Nevertheless formulas \eqref{E127}, \eqref{E5} and \eqref{E6} are changed accordingly so that the kernel is truly the same as in \cite{AJvMTacnode}. }
\bea
p(\Delta \tau;\xi_1,\xi_2) &\&:=\frac{e^{-\frac{(\xi_1-\xi_2)^2}{4\Delta \tau }}}{\sqrt{4\pi \Delta \tau }} \\
\Ai^{(\tau)}(x)&\&:=
\frac{2^\frac 1  6 }{2\pi i}\int _{\gamma_R }  dz~e^{ z^3/3+z^2\tau- zx }=
2^\frac 1  6e^{ \tau x+\frac 2 3\tau ^3 }\Ai (x+\tau^2),
\\
K_{ \Ai }^{(\tau_1,-\tau_2)}(x,y)&\&:=\int_0^{\infty}\!\!
\Ai^{(\tau_1)}(x\!+\!u \sqrt[3]{2})  \Ai^{(-\tau_2)}(y\!+\!u \sqrt[3]{2})du, \label{E5}
\\
{\cal A}^{\tau}_\xi(u) 
&\& := \Ai^{(\tau)} ( \xi+2^{1/3} u)- \int_0^\infty \Ai^{(\tau)} (-\xi+2^{1/3} v )
\Ai (u+ v )dv  ,  \label{E6}
\eea
\ed 
%

Given $r$ times $\tau_1<\tau_2<\dots <\tau_r$ we associate to each a copy of $\R$ or, equivalently we consider the Cartesian product $\R\times \{\tau_1,\dots, \tau_r\}$ which is set-theoretically isomorphic to $\R^r$. Another convenient way of thinking about this set is as the {\em disjoint union} of $\R$ with itself, $r$ times. We will use the notation $\R_{\tau_j}$ to refer to the $j$-th copy (which should be thought of as a copy of $\R$ ``at time $\tau_j$''). 
In each copy $\R_{\tau_j}$ we consider a Borel bounded subset $E^{(j)}$; although it is much less general, the reader should imagine that each $E^{(j)}$ is a finite union of bounded intervals as follows
\be
E^{(j)} = [a_1^{(j)}, b_1^{(j)}] \cup \dots \cup  [a_{\ell_j} ^{(j)}, b_{\ell_j} ^{(j)}]\ ,\ \ a_1^{(j)}<b_1^{(j)}<a_2^{(j)}<b_2^{(j)}<\dots<a_{\ell_j}^{(j)}<b_{\ell_j}^{(j)}.
\ee
We denote by $\mathcal T_\s(\tau)$ the tacnode field at time $\tau$; the {\em joint gap probability} of the multi-set $E:=E^{(1)} \sqcup \dots \sqcup E^{(r)}$ is the probability that there are no points of the tacnode process in $E^{(j)}$ at time $\tau_j$, for all $j=1,\dots, r$: 
\bea
F_{\mathrm{tac}}(E, \vec \tau;\s) := {\rm Prob} \le\{
\mathcal T_\s(\tau_j) \not \in E^{(j)}\ ,\ \ j=1,\dots,r
\ri\}.
\eea
Then the general theory of determinantal random point processes states that 
\bea
F_{\mathrm{tac}}(E, \vec \tau;\s) =  \det \le( \1 - \mathbb K^{\rm tac}\bigg|_E\ri)\label{gaptac}
\eea
where the operator $\mathbb K^{\rm tac}$ is intended as the operator on $L^2(\R_{\tau_1} \sqcup \dots \sqcup \R_{\tau_r})$ with kernel as in Definition \ref{defKtac}.
The main goal is to express the determinant \eqref{gaptac} as the {\em ratio} of two simpler Fredholm determinants. The following two theorems, nevertheless, will be stated for the (more general) {\em generating functions} of the occupation numbers (see \cite{Sosh:RandomPointFields} for details); if $E^{(i)}$ is a disjoint union of intervals $E^{(j)} = \bigsqcup_{\alpha=1}^{\ell_j}  E^{(j)}_{\alpha}$, this latter is defined by
\be
F_{\rm tac}(E, \vec \tau;\s, \vec z):= \le\langle \prod_{j=1}^r \prod_{\alpha=1}^{\ell_j}   (z^{(j)}_{\alpha})^{\sharp_{E^{(j)}_\alpha}}\ri\rangle=
\det \le[
\1 - 
\le(\sum_{j=1}^r \sum_{\alpha=1}^{\ell_j} (1-z^{(j)}_\alpha) \Pi_{E^{(j)}_{\alpha}} \ri) \mathbb K^{\rm tac} \vec \Pi 
\ri]\label{tacgen}
\ee
where $\vec \Pi = \sum_{j=1}^r \sum_{\alpha=1}^{\ell_j} \Pi_{E^{(j)}_{\alpha}}$ is the projector on the multi-interval $\vec E$ and the generating function is a series on the variables $z_\a^{(j)}$. The gap probability is nothing but the value of the generating function at $z^{(j)}_\a=0$.
We shall use the notations 
\be
\Pi^{(j)}_z:=  \sum_{\alpha=1}^{\ell_j} (1-z^{(j)}_\alpha) \Pi_{E^{(j)}_{\alpha}}\ ,\ \ \ 
\vec \Pi_z:= \sum_{j=1}^r\Pi^{(j)}_z.
\ee

\begin{theorem}
\label{P1} 
The generating function \eqref{tacgen} admits the equivalent representation
\bea
F_{\rm tac}(E, \tau;\s,\vec z) =  F_2(\wt \s)^{-1}\!\det \le( \1  - \le[ \begin{array}{c|ccc}
 \pi K_\Ai\pi & -\pi \mathfrak A_{-\tau_1}^T\Pi^{(1)} & \dots &  -\pi \mathfrak A_{-\tau_r}^T\Pi^{(r)} \\
 \hline
 -\Pi^{(1)}_z \mathfrak A_{\tau_1} \pi &  
 \\
 \vdots &&\vec \Pi_z\le( \mathbb K_0 - \mathbb G\ri)\vec \Pi&\\
  -\Pi^{(r)}_z \mathfrak A_{\tau_r} \pi & 
\end{array}\ri]\ri) \nonumber\\
\label{explode0}
\eea
where $F_2(\wt \s)$ is the Tracy--Widom distribution \eqref{TWd}
and \\ $\mathbb K_0, \ \mathbb G: L^2(\R_{\tau_1}  \sqcup \dots \sqcup \R_{\tau_r} ) \to L^2(\R_{\tau_1}  \sqcup \dots \sqcup \R_{\tau_r} ) $ are the operators with kernels
\bea
(\mathbb K_0)_{ij}(\xi_1,\xi_2) =K_\Ai^{(\tau_i, -\tau_j)}(\s-\xi_1, \s-\xi_2)\ ,\qquad 
\mathbb G_{ij}(\xi_1,\xi_2)  = \1_{\tau_1>\tau_2}
\frac{e^{-\frac{(\xi_1-\xi_2)^2}{4(\tau_1-\tau_2) }}}{\sqrt{4\pi (\tau_1-\tau_2) }}
\eea
while $\mathfrak A_{\vec \tau}: L^2(\R)\to L^2(\R_{\tau_1}  \sqcup \dots \sqcup \R_{\tau_r} )  $ has kernel 
\be
[\mathfrak A_{\vec \tau}(x,u)]_{j} =  \Ai^{(\tau_j)} ( x-\s+2^{1/3} u)- \int_0^\infty \Ai^{(\tau_j)} (\s-x+2^{1/3} v )
\Ai (v+u )dv.
\ee
\end{theorem}
{\bf Proof}.
With the notations introduced above, the tacnode kernel defines an operator on\\ $L^2(\R_{\tau_1}  \sqcup \dots \sqcup \R_{\tau_r} ) $ that equals to $[\mathbb K_0 - \mathbb G]_{ij}+ \mathfrak A_{\tau_i}   (\1 -\pi  K_\Ai \pi )^{-1} \mathfrak A_{-\tau_j}^T $.\\
The identity is based on the following operator identity (all being trace-class perturbations of the identity)
\bea
&& \det \le(\1  - \le[ \begin{array}{c|c}
 \pi K_\Ai \pi & -\pi \mathfrak A_{-\vec \tau}^T\vec \Pi\\
 \hline
 -\vec \Pi_z \mathfrak A_{\vec \tau} \pi &\vec \Pi_z\le( \mathbb K_0 - \mathbb G\ri)\vec \Pi
\end{array}\ri]\ri) =\cr
&&=
 \det\le[\begin{array}{c|c}
 \1 -\pi K_\Ai \pi  & 0 \\ 
 \hline 
 0&\1 \end{array}
 \ri] \det  \le[
\begin{array}{c|c}
\1 & 0\\
\hline
\vec \Pi_z \mathfrak A_{\vec \tau }\pi & \1
\end{array}
\ri]
\det \le[ \begin{array}{c|c}
 \1 &(\1 -\pi  K_\Ai \pi )^{-1}\mathfrak A_{-\vec \tau}^T\vec \Pi\\
 \hline
 0 & \1 - \vec \Pi_z\le( \mathbb K_0 - \mathbb G + \mathfrak A_{\vec \tau}  (\1 -\pi  K_\Ai \pi )^{-1}\mathfrak A_{-\vec \tau}^T\ri)\vec \Pi
\end{array}\ri] = \cr
&&=\underbrace{\det \le(\1- \pi K_\Ai \pi     \ri)}_{F_2(\wt \s)} \det( \1 -\vec  \Pi_z {\mathbb K}^{\rm tac} \vec \Pi) 
\eea
\QED

\br[Tacnode gap probabilities as (formal) conditioned process]
The gap probabilities of the tacnode process are  the ratio of two Fredholm determinants: the denominator is the  Tracy-Widom distribution, i.e., the gap probability of the Airy process. It is enticing to interpret thus the ratio as a {\em conditional probability}.  For this interpretation to hold the numerator of formula \eqref{explode0} should be a gap probability of a determinantal process on the configuration space $\R_0 \sqcup \left(\bigsqcup _{j=1}^r \R_{\tau_j}\right)$ with kernel as indicated in the formula itself. 

Then the tacnode would be the {\bf conditioned process} where the conditioning is that there are no points in $[\wt \s,\infty)\subset \R_0$. This type of of conditioned processes have been already analyzed in \cite{Lyons} for the discrete case, and in \cite{TaoWigner} for the continuous case with a finite number of particles. In Appendix \ref{condproc} we treat the continuous case with an arbitrary number of particles and give an explicit expression for the kernel of the conditioned process. However, the correlation functions of this ``extended'' tacnode process fail the positivity test (we checked on some numerical example) and thus the interpretation is only formal.
\er
The second theorem is an equivalent representation where the operator in the  Fredholm determinant in the numerator of Theorem \ref{P1} is further ``unraveled'' in terms of elementary terms.
\begin{theorem}
\label{main}
With the same notations as in Theorem \ref{P1} we have  
\be
F_{\rm tac}(E, \vec \tau;\s, \vec z)= 
\frac {\det \le[ \1 -\wh  \Pi_z  \mathbb H\wh \Pi \ri]       }{\det \le[\1 -\pi K_{\Ai} \pi \ri]}
\ee
where the Fredholm determinant is defined on the space $$
L^2(\R_+ \sqcup \R_0 \sqcup \R_{\tau_1} \sqcup\dots, \sqcup \R_{\tau_r}) \simeq 
L^2(\R_+)\oplus L^2( \R_0)\oplus \left(\R_{\tau_1} \sqcup\dots \sqcup \R_{\tau_r}\ri).$$ 
Denoting the $L^2(\R_+) $ with the index $_{-1}$, the operators that appear are the following ones 
\bea
\wh \Pi_z := \1_{-1} \oplus \pi \oplus \vec \Pi_z\ ,\ \ \wh \Pi = \wh \Pi_0 = \Id_{-1} \oplus \pi \oplus \vec \Pi
\eea 
and their kernel $\mathbb H$ is expressed in terms of ordinary Airy functions as follows
\bea
 \left[
\begin{array} {c|c|c}
  \mathbb H_{-1,-1} \equiv 0
& 
 \mathbb H_{-1,0}(x,y) = -\Ai(x+y) 
&
\!\!  \mathbb H_{-1,j}(x,y) =  \Ai^{(-\tau_j)} (x\sqrt[3]{2}+\s-y  ) \!\!\!
 \\[10pt]
 \hline &&\\ 
 \mathbb H_{0,-1}(x,y) =  - \Ai  (x+y ) 
&
 \mathbb H_{0,0}\equiv 0
 &
   \mathbb H_{0,j}(x,y) = \Ai^{(-\tau_j)} (x\sqrt[3]{2} + y-\s ) 
 \\[10pt]
 \hline &&\\
\!\!\!  \mathbb H_{i,-1}(x,y) = \Ai^{(\tau_i)} (\s-x+y \sqrt[3]{2}  )\!\!
 &
\!  \mathbb H_{i,0}(x,y) = \Ai^{(\tau_i)} (x-\s+y\sqrt[3]{2} ) \!\!
  &
   \mathbb H_{i,j}(x,y) = - p(\tau_i - \tau_j; x,y)\chi_{i >j}
\end{array}
\right],\nonumber\eea
with $1\leq i,j \leq r$.
\end{theorem}
\bigskip

The expression of the Fredholm determinant in Theorem \ref{main} is the most suitable for the numerical computations because each entry of the kernel is either a simple modification of the Airy function or exponential.  The gap probabilities are obtained simply by setting all the $z_\alpha^{(j)}$'s to zero, or --which is the same-- restricting the operator $\mathbb H $ on the set  $\R_+ \sqcup [\wt \s,\infty) \sqcup\left(\bigsqcup_{j=1}^R E^{(j)}\right)$.

{\bf Proof of Theorem \ref{main}.}
For brevity we shall introduce the operators  with the kernels below
\bea
&& \mathcal C_\tau  (\xi,u):=2^\frac 1 6  \Ai^{(\tau)} \le(\xi -\s+ u\sqrt[3]{2}\ri):L^2(\R_+)\to L^2(\R_\tau)\\
&& \mathcal C(u,v):= \Ai(u+v) :L^2(\R_+) \to L^2(\R) 
\eea
The first observation is that both $K_\Ai$ and $K_\Ai^{(\tau_1,-\tau_2)}$ can be written as compositions; 
\be
K_\Ai = \mathcal C \mathcal C^T\ ,\qquad K_{\Ai}^{(\tau_1,-\tau_2)} = \mathcal C_{\tau_1}  \mathcal C_{-\tau_2}^T\ .
\ee
If we denote the parity operator around $\s$ by $P:L^2(\R)\to L^2(\R)$ $(Pf)(x) = f(2\s-x)$ then 
the tacnode kernel can be written as
\be
\mathbb K^{\rm tac}_{ij} =
 P\mathcal C_{\tau_i} \mathcal C_{-\tau_j}^T P 
 + \mathfrak A_{\tau_i} \pi \le(\1 - \pi \mathcal C \mathcal C^T \pi\ri)^{-1}\pi \mathfrak A_{-\tau_j}^T - \mathbb G_{ij}
\ee
where 
\bea
[\mathfrak U_\tau]_j = \mathcal C_{\tau_j} - P\mathcal C_{\tau_j} \mathcal C: L^2(\R_+)\to L^2(\R_{\tau_j})
\eea
%
%
We denote $\mathcal C_{\vec \tau}: L^2(\R_+)\to  L^2(\bigsqcup_{j=1}^r \R_{\tau_j})\simeq \bigoplus_{j=1}^r L^2(\R_{\tau_j})$ as the operator with components $\mathcal C_{\tau_j}$; 
then the proof relies upon
the following identity of determinants
\bea
\det \le(\1  - \le[ \begin{array}{c|c}
 \pi \mathcal C \mathcal C^T \pi & -\pi \mathfrak A_{-\vec \tau}^T\vec \Pi\\
 \hline
 -\vec \Pi_z \mathfrak A_{\vec \tau} \pi &\vec \Pi_z(\mathbb  K_0 - \mathbb G)\vec \Pi
\end{array}\ri]\ri) 
= \det  \le[
\begin{array}{c|c|c}
 \1 &\mathcal C^T\pi  &    - \mathcal C_{-\vec \tau}^T P \vec \Pi \\
 \hline
 \pi \mathcal C & \1 & -\pi\mathcal C_{-\vec \tau}^T \vec \Pi  \\
\hline
 -\vec \Pi_z P \mathcal C_{\vec \tau}  &-\vec \Pi_z \mathcal C_{\vec \tau} \pi  &\1+ \vec \Pi_z \mathbb G \vec \Pi 
\end{array}
\ri] \label{Hc}
\eea
The identity is seen by 
\bea
 \le[
\begin{array}{c|c|c}
 \1 &\mathcal C^T\pi  &    - \mathcal C_{-\vec \tau}^T P \vec \Pi \\
 \hline
 \pi \mathcal C & \1 & -\pi \mathcal C_{-\vec \tau}^T\vec  \Pi  \\
\hline
 -\vec \Pi_z P \mathcal C_{\vec \tau}   &-\vec \Pi_z \mathcal C_{\vec \tau}\pi  & \1+ \vec \Pi_z \mathbb G \vec \Pi 
\end{array}
\ri] 
 \le[
\begin{array}{c|c|c}
 \1 &-\mathcal C^T\pi  &     \mathcal C_{-\vec \tau}^T P \vec \Pi \\
 \hline
0 & \1 & 0  \\
\hline
0  &0 & \1 
\end{array}
\ri] 
= \cr
 \le[
\begin{array}{c|c|c}
 \1 &0 &   0 \\
 \hline
 \pi \mathcal C & \1 -\pi \mathcal C\mathcal C^T\pi & -\mathcal C_{-\vec \tau}^T \vec \Pi + \pi \mathcal C\mathcal C_{-\vec \tau }^TP\vec \Pi   \\
\hline
 -\vec \Pi_z P \mathcal C_{\vec \tau}  &-\vec \Pi_z \mathcal C_{\vec \tau} + \vec \Pi_z P \mathcal C_{\vec \tau} \mathcal C^T\pi  &\1+ \vec\Pi_z \mathbb G \vec \Pi - \vec\Pi_z P \mathcal C_{\vec \tau} \mathcal C_{-\vec \tau}^T P \vec \Pi 
\end{array}
\ri] 
\eea
The operator $\mathbb H$ is read off \eqref{Hc}. 
\QED

\subsection{From the Tacnode to the Pearcey  process}

\begin{figure}
\includegraphics[height=0.4\textwidth,width=0.45\textwidth]{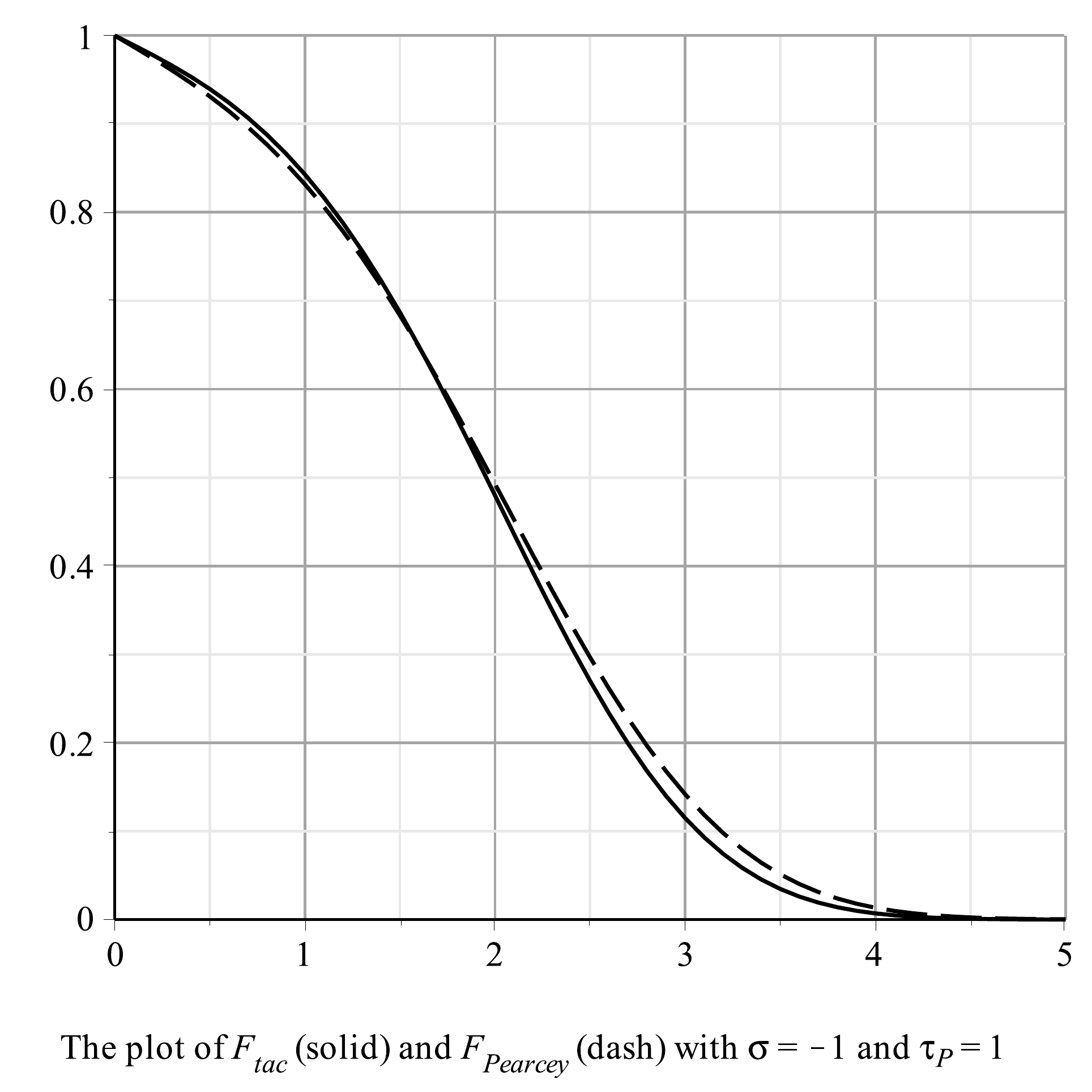}
\includegraphics[height=0.4\textwidth,width=0.45\textwidth]{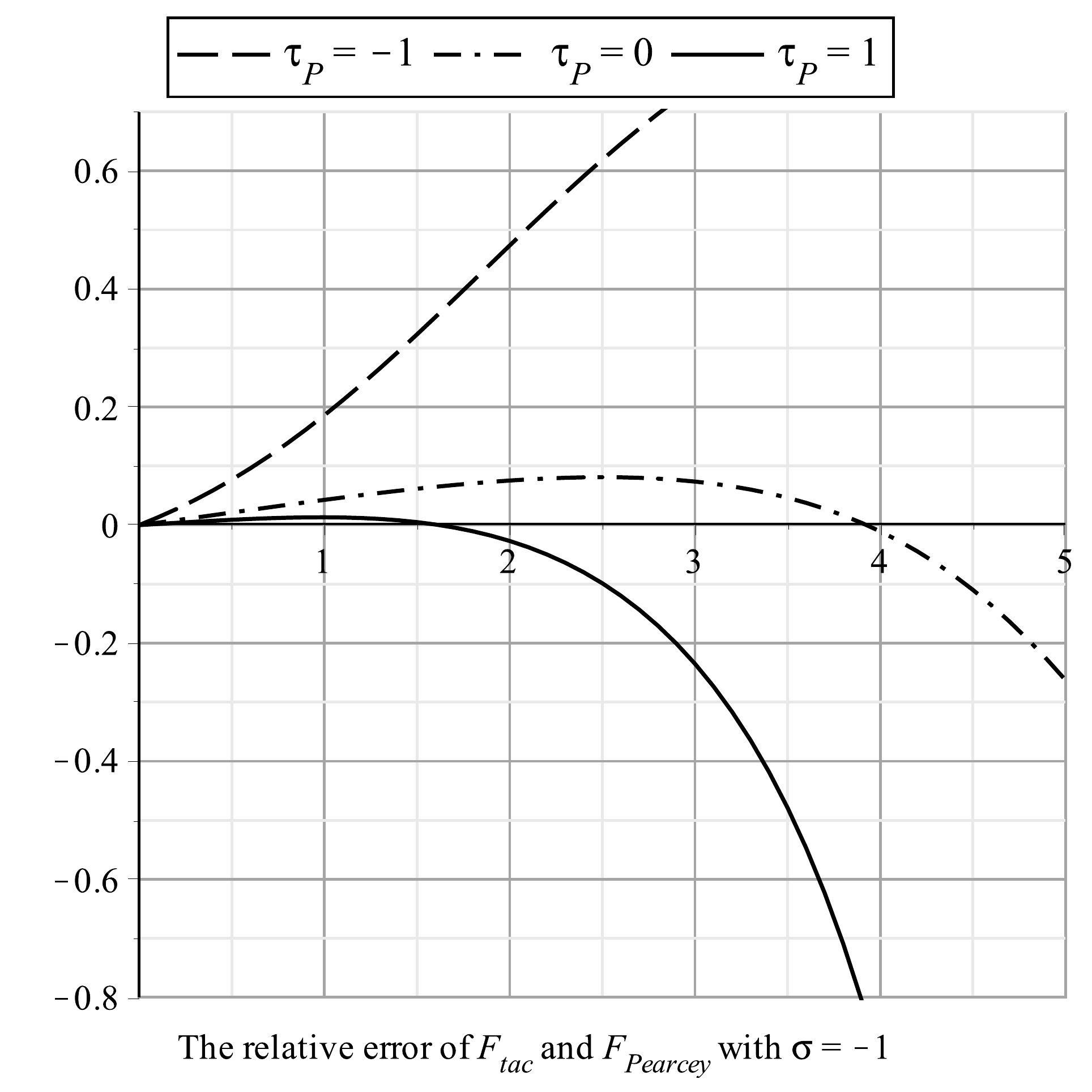}

\includegraphics[height=0.4\textwidth,width=0.45\textwidth]{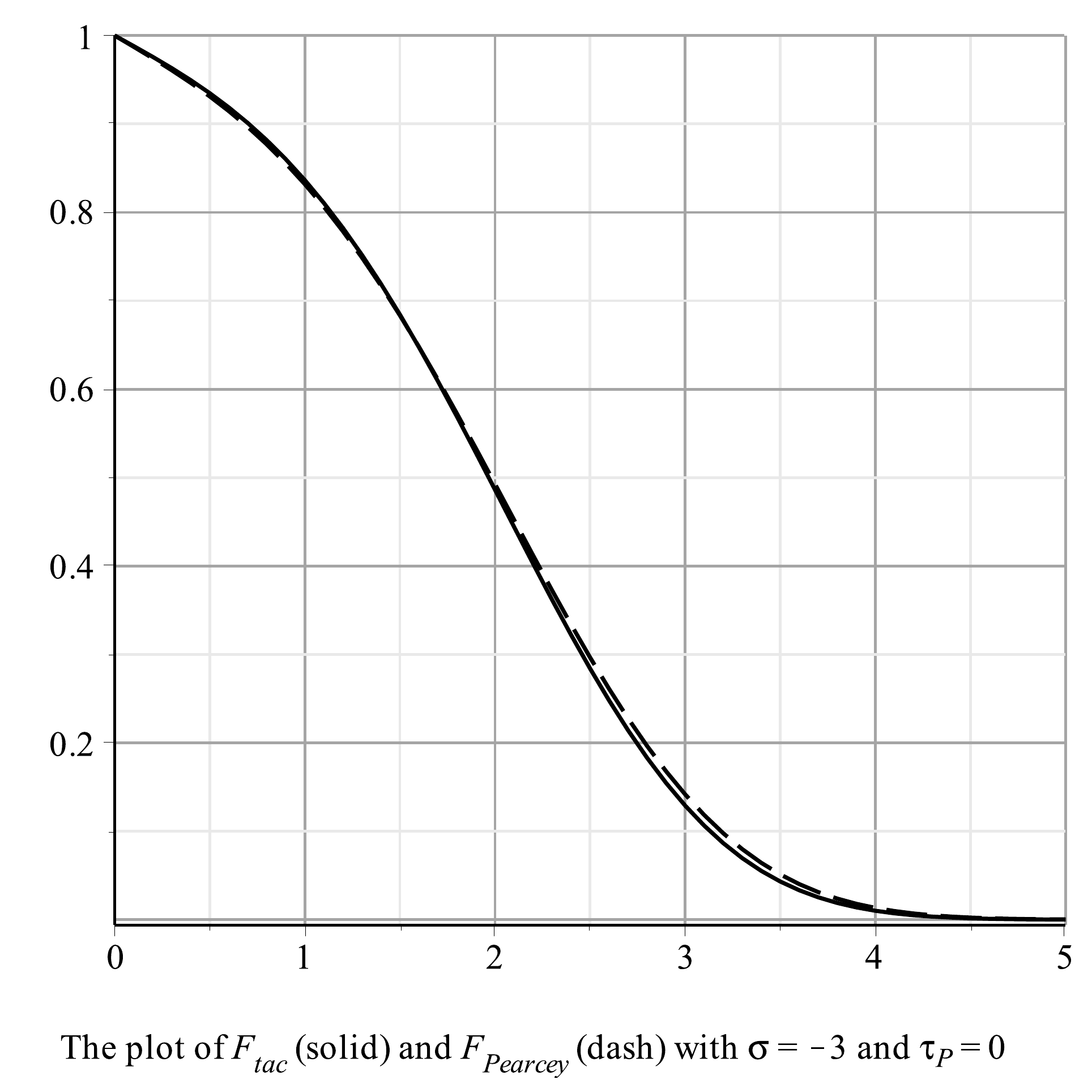} 
\includegraphics[height=0.4\textwidth,width=0.45\textwidth]{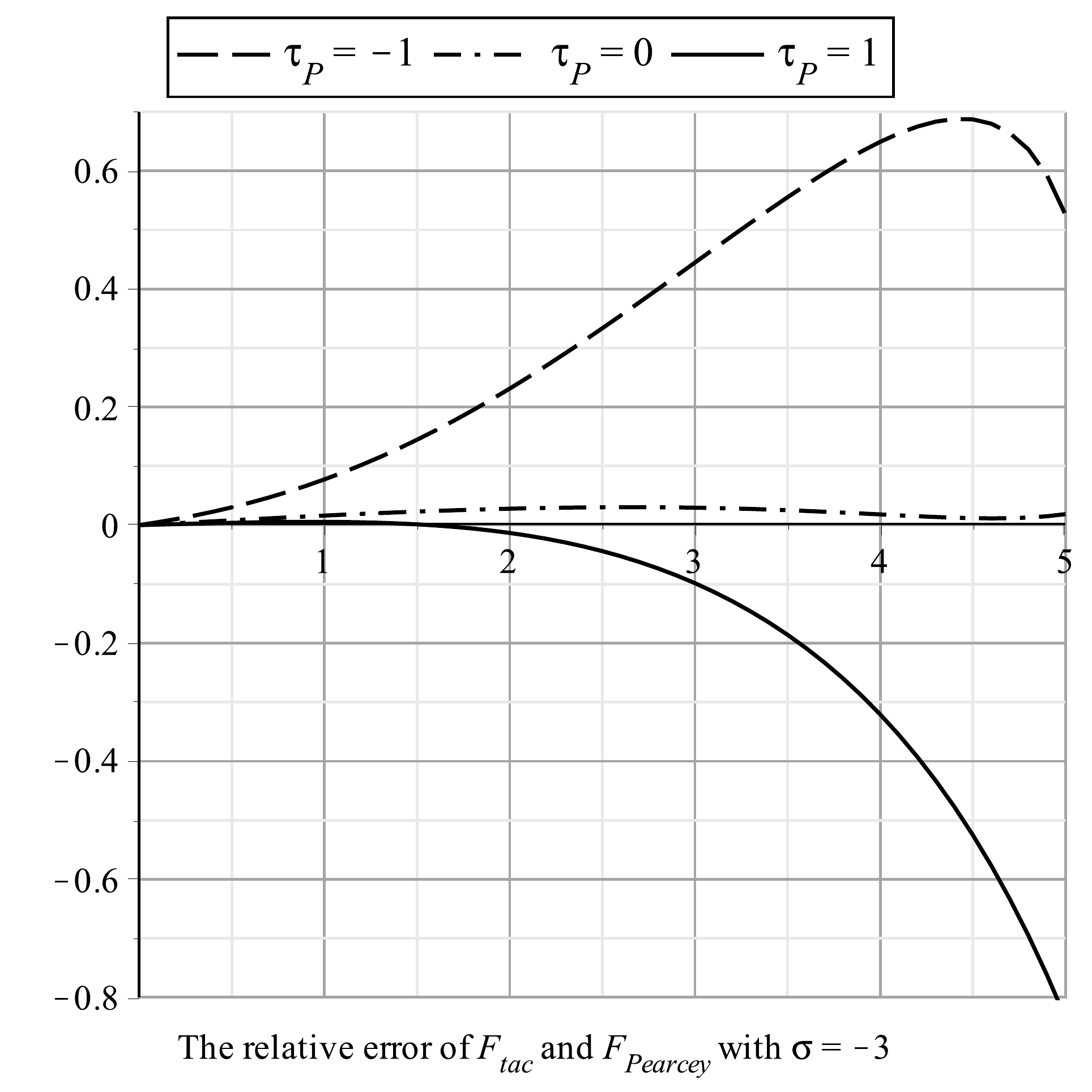}

\includegraphics[height=0.4\textwidth,width=0.45\textwidth]{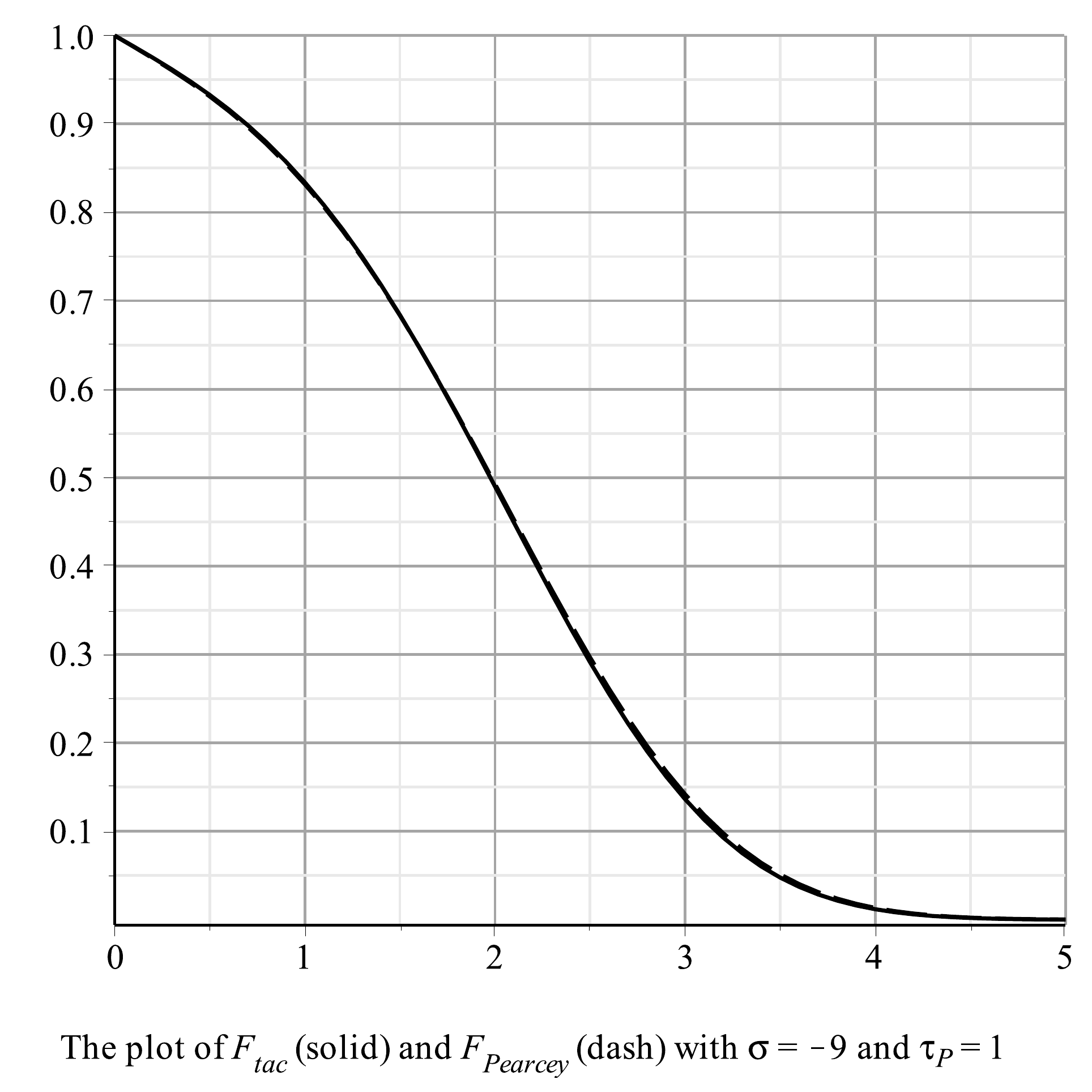} 
\includegraphics[height=0.4\textwidth,width=0.45\textwidth]{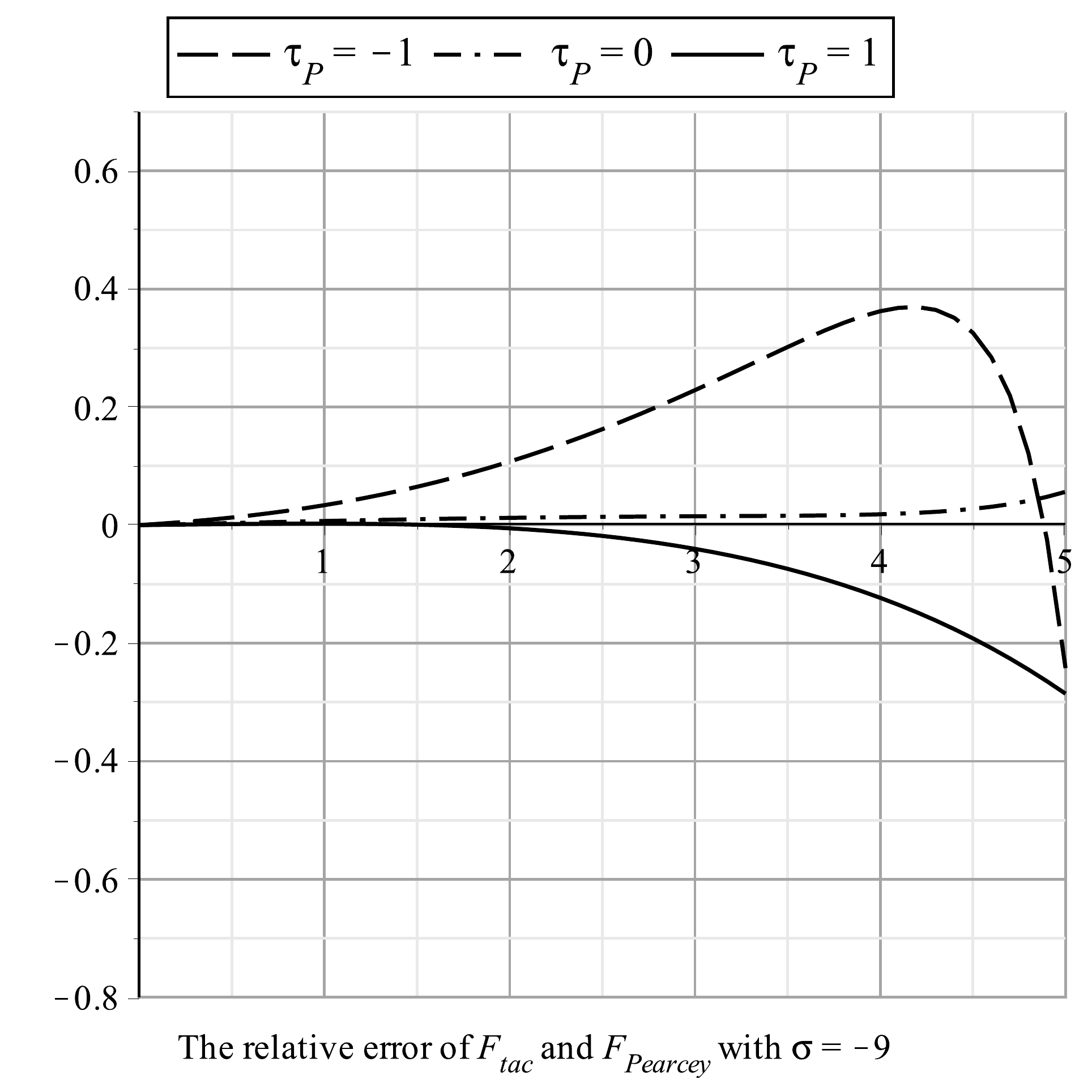}
\caption{\footnotesize The graphs of $F_{\rm tac}([-a_{tac},a_{tac}],\tau_{tac};\s)$ and $F_{\rm P}([-a_P,a_P]; \tau_{P})$  with the parameters related by \eqref{goodone}.}
\label{TactoP1t}
\end{figure}

\begin{figure}
\includegraphics[height=0.4\textwidth,width=0.45\textwidth]{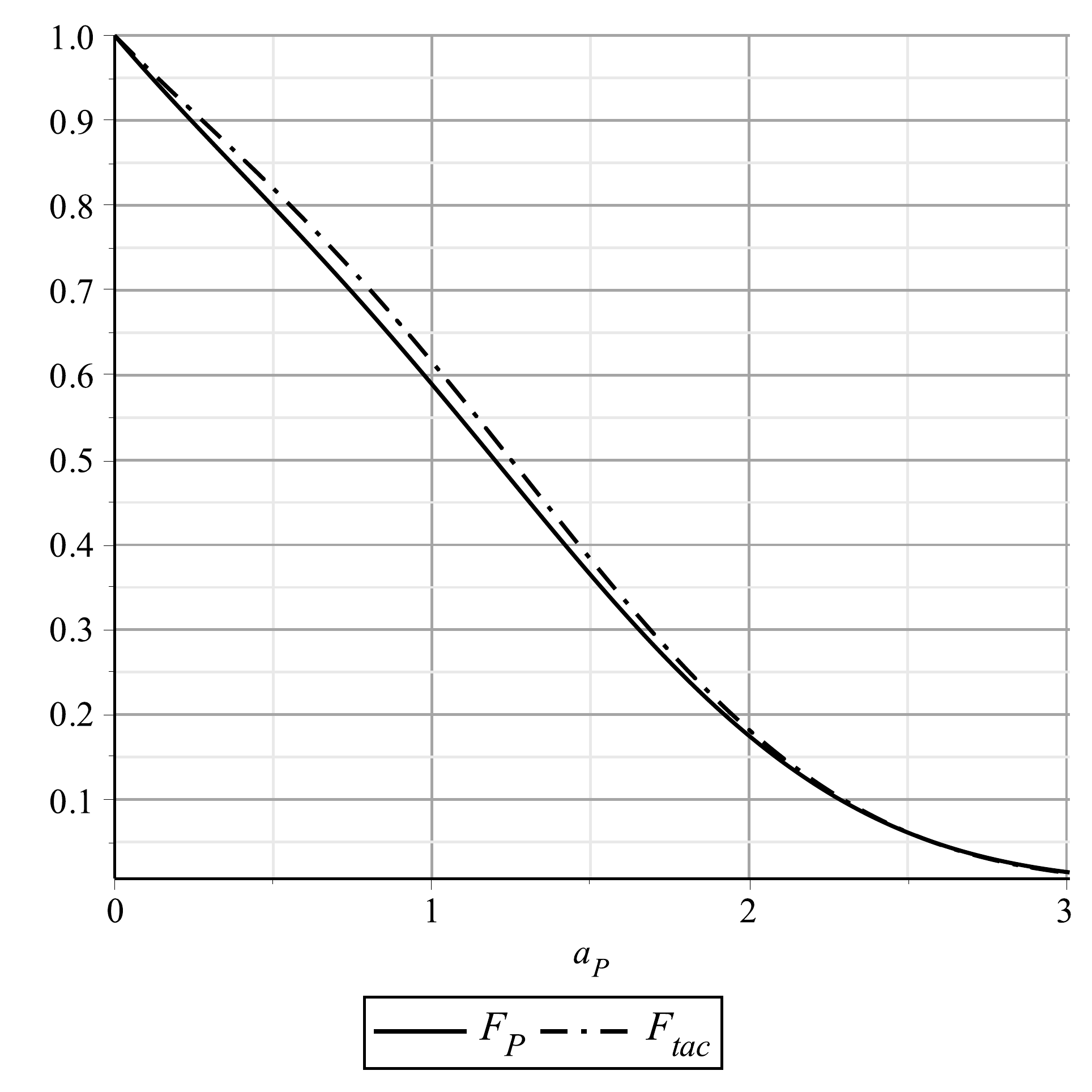}
\includegraphics[height=0.4\textwidth,width=0.45\textwidth]{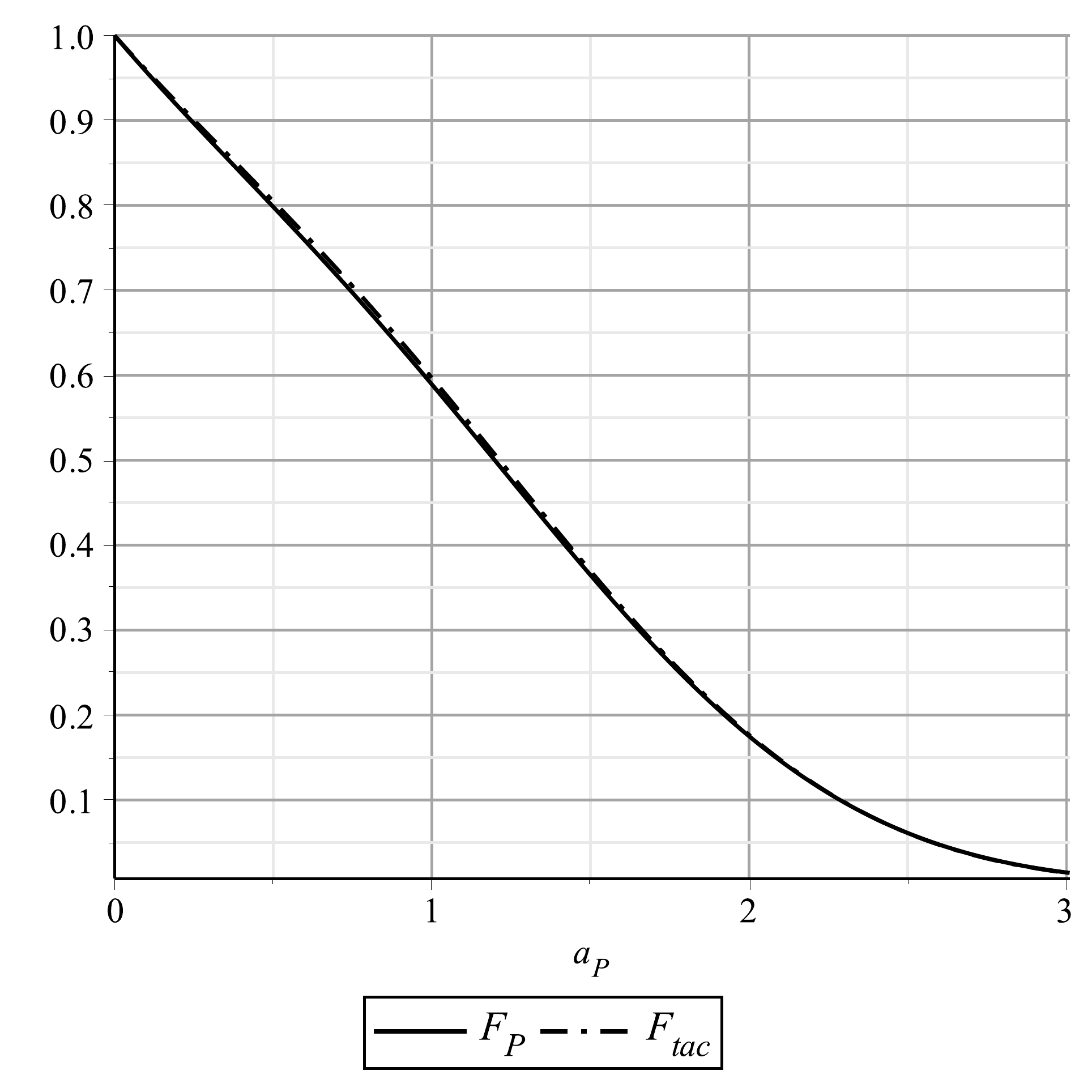} 
\includegraphics[height=0.4\textwidth,width=0.45\textwidth]{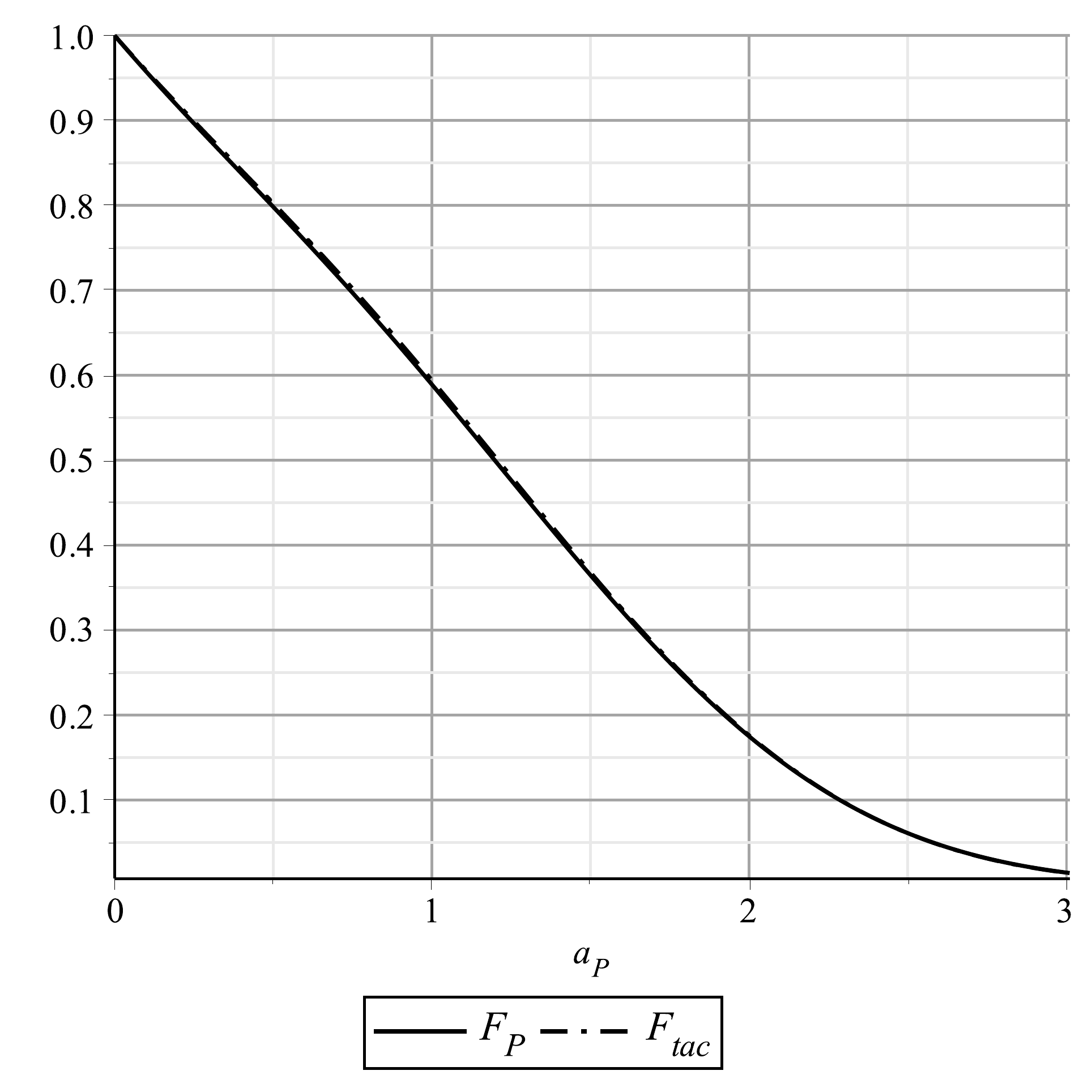} 
\includegraphics[height=0.4\textwidth,width=0.45\textwidth]{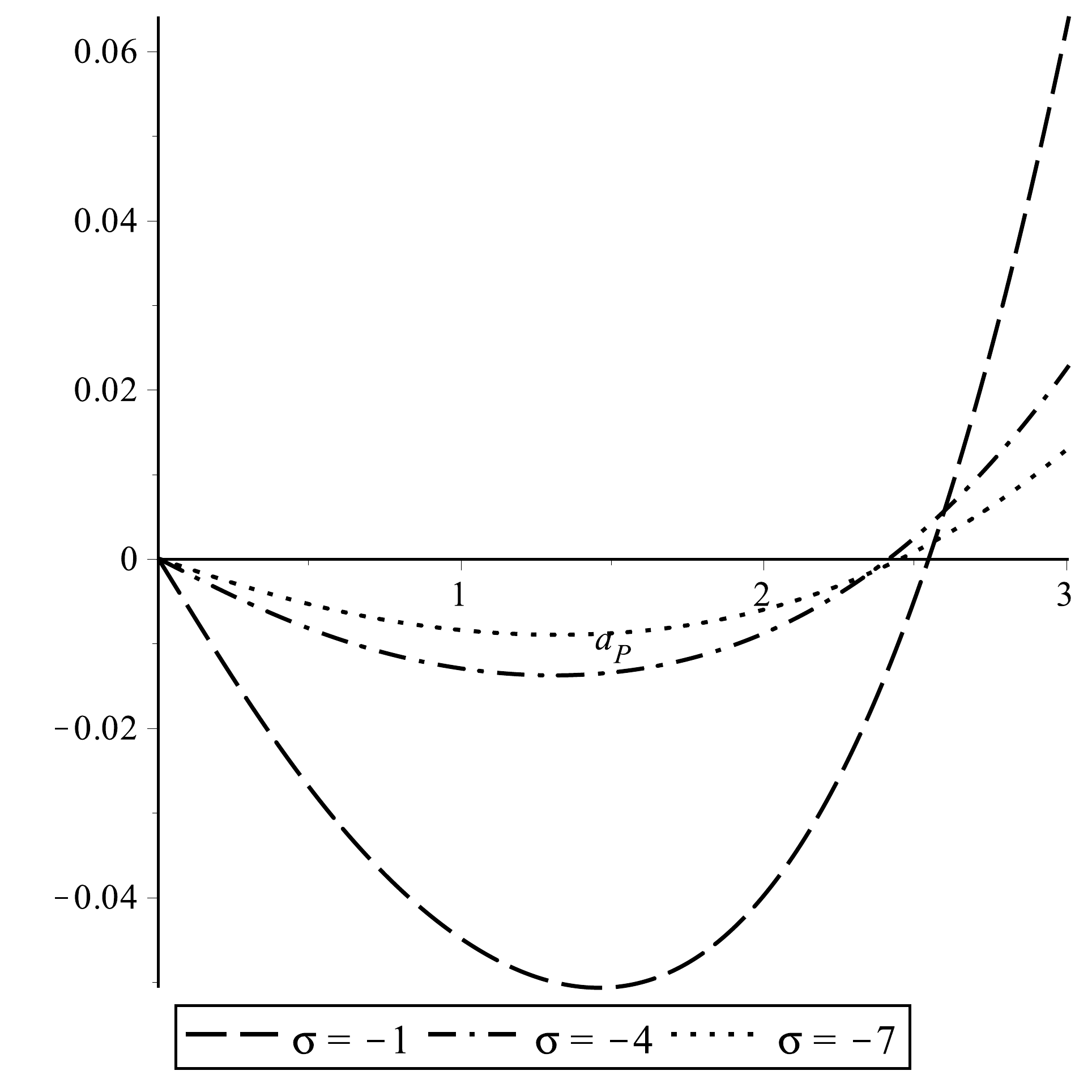}
\caption{\footnotesize The multi--time case. Here the graphs of $F_{\rm tac}\le(\le\{[-a_{tac},a_{tac}]; \tau_{tac,1},[-a_{tac},a_{tac}]; \tau_{tac,2}\ri\};\s\ri)$ and $F_{\rm P}\le(\le\{[-a_P,a_P]; \tau_{P,1},[-a_P,a_P]; \tau_{P,2}\ri\}\ri)$ are shown with the parameters related by \eqref{goodone} and $\tau_{P,1}=0, \tau_{P,2}=1$. The last graph is the plot of the relative discrepancy between $F_{\rm P}$ and $F_{\rm tac}$.}
\label{TactoP2t}
\end{figure}

Given its origins in the Dyson Brownian diffusion, and by the same idea that led to considering the limit of the Pearcey to the Airy process,  it is physically expected that the tacnode process should converge to the Pearcey process under a suitable rescaling sending the overlap $\sigma$ to $-\infty$, which amounts to ``push'' closer and closer  the two sets of particles touching on the tacnode point. 
A rigorous approach based on the representation of the gap-probabilities in terms of a Riemann--Hilbert problem \cite{BCG1} suggests that this convergence occurs in the following asymptotic regime
\be
a_{tac} = \frac {a_{P}} {(-8\s)^\frac 18}\ ,\qquad \tau_{tac} = \pm \sqrt{\frac {-\s}2} + \frac {\tau_P}{(-2^7\s)^{\frac 1 4}}\ ,\ \ \ \s\to -\infty;\label{goodone}
\ee
in other words we have that
\be
	\lim_{\sigma\rightarrow -\infty} F_{\rm tac}\left(\left[\frac {a_{P}} {(-8\s)^\frac 18},\frac {b_{P}} {(-8\s)^\frac 18}\right],\pm\sqrt{\frac {-\s}2} + \frac {\tau_P}{(-2^7\s)^{\frac 1 4}},\s\right) = F_{\rm P}\Big(\big[a_P,b_P\big],\tau_P\Big). \label{goodone2}
\ee
Indeed, modulo a multiplicative rescaling of the variables (but without changing the scale exponents) the formula above is proven in \cite{GeudensZhangTacPearcey}, Theorem 2.3 (more correctly, \eqref{goodone2} is a direct consequence of the quoted theorem, since the uniform convergence of the kernels on compact sets implies the convergence of the Fredholm determinants). In \cite{BCG1} we will give another proof of \eqref{goodone2}, based on the nonlinear steepest descent analysis of a \emph{different} Riemann-Hilbert problem giving directly the gap probability of the process rather than its kernel.\\
Here we show numerically the validity of \eqref{goodone2} by computing the following two gap-probabilities:
\bea
F_{\rm tac} ([a_{tac} ,b_{tac}]; \tau_{tac}, \s) := \mathbb P_\s\le\{\mathcal T_\s(\tau_{tac}) \not\in[a_{tac} , b_{tac}] \ri\}\\
 F_{\rm P} ([a_{P}, b_P], \tau_{tac}) :=  \mathbb P\le\{\mathcal P(\tau_{P}) \not\in[-a_{P} , a_{P}] \ri\}
\eea
and verifying numerically the convergence. This has been done for $-\s$ up to $9$, after which our\footnote{None of the author specializes in numerical computations, and hence we have implemented the ideas of \cite{Bornemann} in its simplest form. See Appendix.} numerical implementation of the Fredholm determinant becomes too unstable, see Fig. \ref{TactoP1t}.
We have also verified in a numerical example the convergence of a two-time gap probability, see Fig. \ref{TactoP2t}.

\subsection{From the tacnode to the Airy process}
There are two types of regimes in which the gap probability for the tacnode process ``degenerates'' into the Airy one:
\begin{enumerate}
\item large separation $\s\to+\infty$; physically it corresponds to pull far apart the two sets of Brownian particles touching on the tacnode point.
\item large time $\tau \to \pm \infty$; it corresponds to move far away from the singular point along the boundary of the space-time region swept out by the particles.
\end{enumerate}
Let us for simplicity describe the one-interval and one-time case. Keeping the overlap $\s$ fixed, we will have that
\be
	\lim_{\tau\rightarrow\infty}F_{\rm tac}\left(\left[a-\s-\tau^2,b-\s-\tau^2\right],\tau,\s\right) = F_2\Big(\big[a,b\big]\Big)
\label{conAi1}\ee
while, keeping $\tau$ fixed, in a complete analogous way we obtain
\be
	\lim_{\s\rightarrow\infty}F_{\rm tac}\left(\left[a-\s-\tau^2,b-\s-\tau^2\right],\tau,\s\right) = F_2\Big(\big[a,b\big]\Big).
\label{conAi2}\ee
This convergence is simple to see by inspecting directly the kernel of the (extended) tacnode process \eqref{E127} because  the term involving the resolvent of the Airy kernel tends to zero, uniformly over compact sets of the translated spatial variables $x-\s - \tau^2$. Here we will not study the rate of convergence in formulas \eqref{conAi1},\eqref{conAi2}. We simply point out that, if we knew some equation for the tacnode gap probability, we could perform a similar analysis to the one in \cite{ACvM2}. The extension to multi--time case (and multi--intervals), though more cumbersome, does not present any additional difficulty.

\begin{figure}
\includegraphics[width=0.49\textwidth]{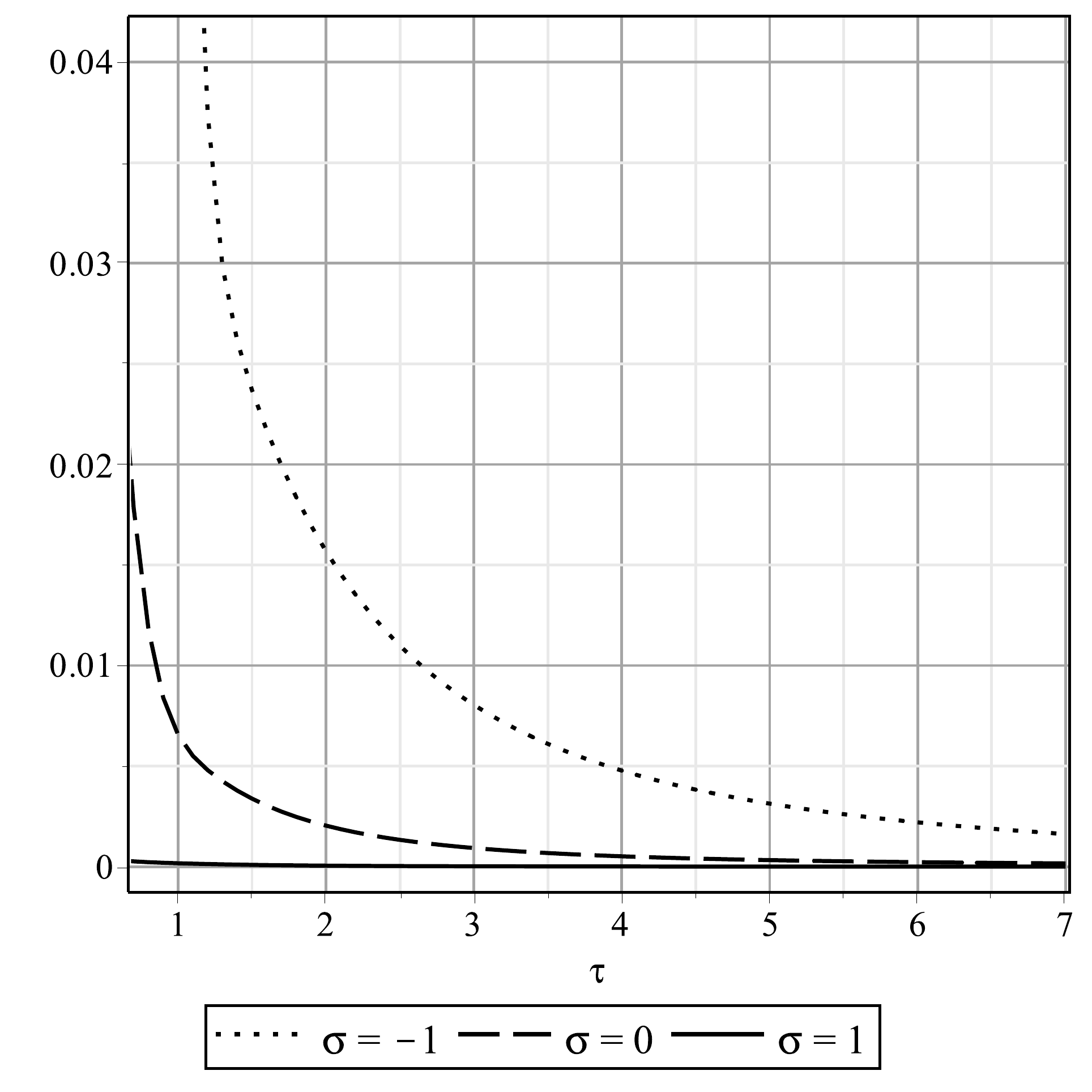}
\includegraphics[width=0.49\textwidth]{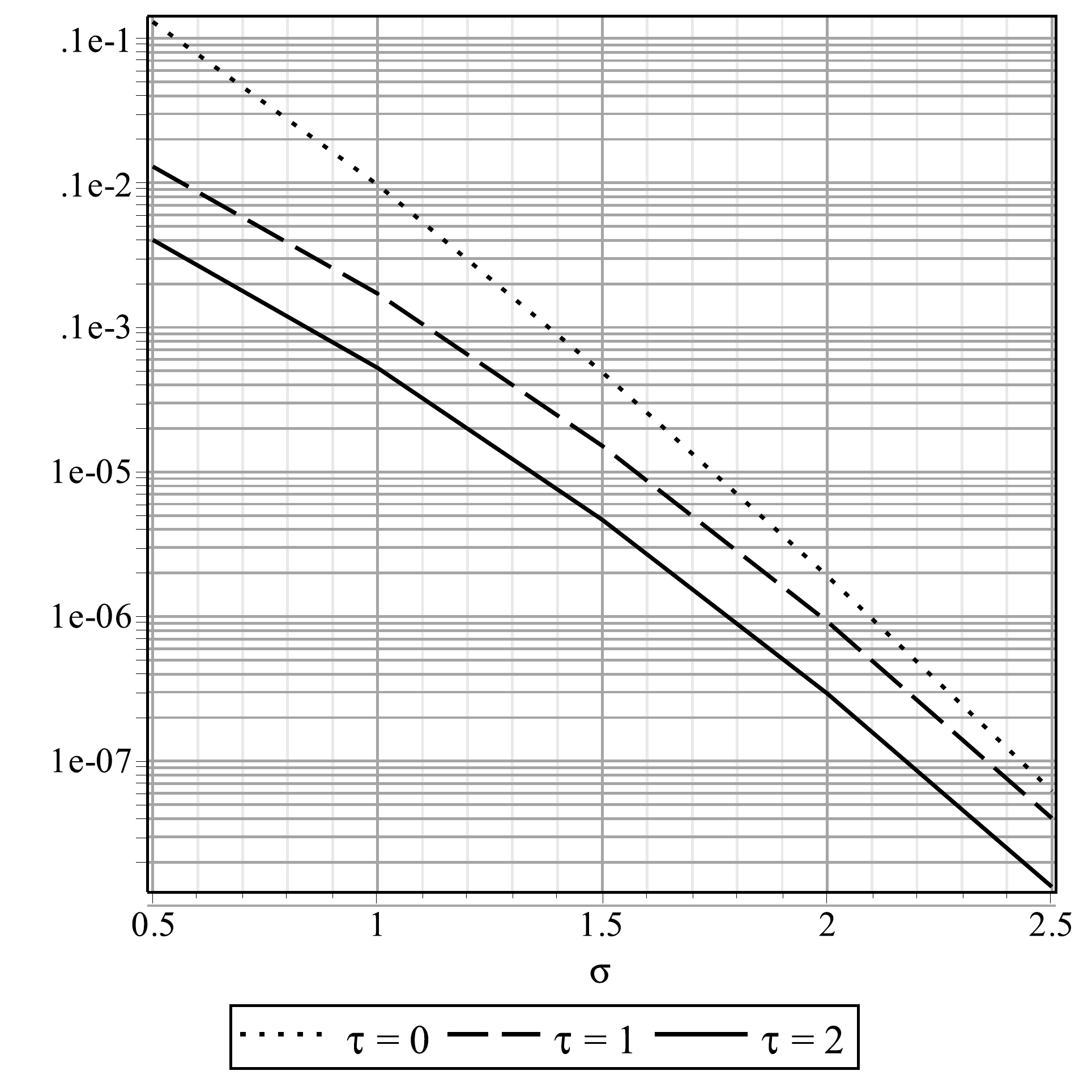}
\caption{\footnotesize The relative values $1 - \frac {F_{\rm tac}([ a_{tac}, b_{tac}]; \tau_{tac}, \s)}{F_2(a)F_2(b)} $ with $a_{tac} =  a-\s-\tau_{tac}^2,\ b_{tac} = b + \s + \tau_{tac}^2$, plotted against $\tau_{tac}$ (left) and $\sigma$ (right), showing the convergence of the tacnode gap probability to the product of two Tracy-Widom distributions. Here, by the way of example,  $a = -0.3, b = 0.5$. The convergence as $\s\to\infty$ is clearly exponential (note that the graph on the right is on a semilog coordinate system).}
\label{TactoAiry}
\end{figure}

A perhaps more interesting situation, less obvious from the formula \eqref{E127} but equally natural from the ``physical'' setting, is the one in which the tacnode process degenerates into a \emph{couple} of Tracy--Widom distributions, in analogy with the Pearcey-to-Airy transition \eqref{PtoAi} proved in \cite{BertolaCafasso1}. In this case, roughly speaking, half of the space variables (endpoints of the gaps) moves far away from the tacnode following the left branch of the boundary of the space--time region swept by the particles, and the other half goes in the opposite direction. The simplest instance of this degeneration is the case of one-time and one-interval $E = [a -\s-\tau^2, -b+\s+\tau^2]:$ 
\be
	\lim_{\s\rightarrow\infty} F_{\rm tac}\Big(\big[a-\s-\tau^2,-b+\s+\tau^2\big],\tau,\sigma\Big) = F_2([a,\infty))F_2([b,\infty)).
\label{112}\ee
Similarly, the limit \eqref{112} holds also  for $\tau\to \pm\infty$  and $\s$ fixed. The formula \eqref{112}, as well as its generalization to the multi--interval case, will be proven in \cite{BCG1}, numerically these regimes are illustrated in Figure \ref{TactoAiry}.

\FloatBarrier

\appendix

%
%

\subsection*{Acknowledgements} 

The work of M.B. is partially supported by the Natural Sciences and Engineering Research Council of Canada (NSERC) and the work of M.C. is partially supported by the ANR project DIADEMS. M.C. thanks the UMI CRM/CNRS for the hospitality and the financial support during his stay in Montreal, and he is also grateful to A.Hardy and L.Chaumont for very useful suggestions and discussions.

\section{Conditioned processes}
\label{condproc}
Consider a  determinantal random point field  (process, DRPF) $\mathcal R$ on the configuration space $\mathcal X$, with kernel $K: \mathcal X \times \mathcal X \to \R$; we assume that $\mathcal X$ is endowed with a  measure $\d\mu$  and $K$ can be interpreted as an operator in $L^2(\mathcal X, \d \mu)$.
The knowledge of the occupation probabilities of any point process  is equivalent to the knowledge of their correlation functions by means of the generating functions of the occupation numbers:
\be
F (E;\vec z):= \le<\prod_{j=1}^k (z_j)^\sharp_{E_j} \ri> = \sum_{\vec \ell \in \mathbb N^k}  \mathbb P \le\{\sharp_{E_j} = \ell_j\ ,\ \ j=1, \dots k\ri\} \prod_{j=1}^k z_j^{\ell_j} 
\ee
where we denote with $E:=E_1 \sqcup \dots \sqcup E_r$ an arbitrary disjoint union of intervals.
Denoting the indicator function of a set by $\Pi_{E}$, also thought of as orthogonal projection, the above generating function is computed as a Fredholm determinant (see \cite{Sosh:RandomPointFields} for details)
\be
F(E;\vec z) = \det \le[\1_{\mathcal X}- \le( \sum_{j=1}^k(1- z_j) \Pi_{_{E_j}}\ri) K \Pi_{_E}\ri].
\label{freddet}
\ee
Let $\mathcal A \subset \mathcal X$. We want to study the conditional process (field) $\mathcal R_{\mathcal A}$ on $\mathcal Y:= \mathcal X \setminus \mathcal A$ as the process ``conditioned to have no points in $\mathcal A$''.
\begin{remark}
For  any $\mathcal A\subset \mathcal X$ denote by $\sharp_{\mathcal A}$ the integer-valued random variable counting the points of a configuration that  belong to $\mathcal A$. 
Let $E = E_1\sqcup \dots \sqcup E_k\subset \mathcal Y$: then the conditional probabilities of occupation numbers $\mathbb P_{\mathcal A}$ are given as usual as
\be\label{conditionals}
\mathbb P_{\mathcal A}\le\{\sharp_{E_j} = k_j\ ,\ \ j=1, \dots k\ri\} := \frac {\mathbb P\le\{\sharp_{E_j} = k_j\ ,\ \ j=1, \dots k, \sharp_{\mathcal A} =0 \ri\}}{\mathbb P\le\{\sharp_{\mathcal A} =0 \ri\}}
\ee
\end{remark}
On the other hand, according to the general setup for determinantal random point fields, the conditional process  is completely determined by all possible generating functions for the occupation numbers, denoted below with $F_{\mathcal A}(E;\vec z)$:

\bea
F_{\mathcal A} (E;\vec z)  = \sum_{\vec \ell \in \mathbb N^k}  \mathbb P_{\mathcal A} \le\{\sharp_{E_j} = \ell_j\ ,\ \ j=1, \dots k\ri\} \prod_{j=1}^k z_j^{\ell_j}  =\\=
  \sum_{\vec \ell \in \mathbb N^k}  \frac {\mathbb P\le\{\sharp_{E_j} = k_j\ ,\ \ j=1, \dots k, \sharp_{\mathcal A} =0 \ri\}}{\mathbb P\le\{\sharp_{\mathcal A} =0 \ri\}}\prod_{j=1}^k z_j^{\ell_j} = \frac{F(E\sqcup \mathcal A;(\vec z,0))}{F(\mathcal A;0)}
  \label{condoccprop}
\eea
where the denominator of the last formula should be understood as
\be
F(E\sqcup \mathcal A;(\vec z,0)) = \det \le[\1_{\mathcal X}- \le(\left( \sum_{j=1}^k(1- z_j) \Pi_{_{E_j}}\right)+\Pi_{\mathcal A}\ri) K \Pi_{E\sqcup \mathcal A}\ri].
\label{freddet2}
\ee
Note, in particular, that both the numerator and denominator are Fredholm determinants of kernels, exactly as in Section 3 above,Theorem \ref{main}. Below we observe that, given a determinantal process, the associated conditional processes are also determinantal themselves (see also \cite{TaoWigner} and \cite{Lyons}).
\begin{prop}
Suppose you are given a determinantal random point field $\mathcal R$ on $\mathcal X$ with kernel $K$, and let $\mathcal A\subseteq \mathcal X$ an arbitrary measurable subset of the state space. Then the associated conditional process $\mathcal R_{\mathcal A}$ on $\mathcal Y:= \mathcal X \setminus \mathcal A$ is also determinantal with kernel 
\be
K_{\mathcal A}: \mathcal Y\times \mathcal Y\to\R
\ee
given by 
\be
K_{\mathcal A}(y_1,y_2)  = K(y_1,y_2) + \int_{\mathcal A} K(y_1,a_1) \le(\1_{\mathcal A} - K\big|_{\mathcal A} \ri)^{-1}(a_1,a_2) K(a_2,y_2) \d\mu(a_1)\d\mu(a_2)
\ee
or, in operator notation,
\be
K_{\mathcal A}:= K + K\Pi_A  \le(\1_{\mathcal A} - K\big|_{\mathcal A} \ri)^{-1}\Pi_{\mathcal A} K\ .
\ee
Here $\Pi_{\mathcal A}$ denotes the projection onto $L^2(\mathcal A)\subset L^2(\mathcal X)$.
\end{prop}
{\bf Proof}
The proof relies on the computation of all possible generating functions. 
We start  the numerator of the right side of  \eqref{condoccprop}: we realize the integral operator defined by $K$ restricted to $E\sqcup \mathcal A = E_1\sqcup \dots \sqcup E_k \sqcup \mathcal A$ as an operator on $L^2(E_1) \oplus\dots \oplus L^2(E_k) \oplus L^2(\mathcal A)$. With obvious block-matrix notation and denoting $\l_j= 1-z_j$, we then have the specialization of the general determinantal formula \eqref{freddet} is  
\bea
F(E\sqcup\mathcal A;(\vec z,0)) = \det \le[ \begin{array}{c|c|c|c}
\1_{\mathcal A} - K_{\mathcal A\mathcal A} &- K_{\mathcal AE_1} &\dots & -K_{\mathcal AE_k}\\
-\l_1K_{E_1\mathcal A} & \1_{E_1} -\l_1 K_{E_1E_1} & \dots &-\l_1 K_{E_1E_k}\\
\vdots &&&\\
-\l_k K_{E_k\mathcal A} & -\l_kK_{E_k E_1} &\dots & \1_{E_k} - \l_k K_{E_kE_k}
\end{array}\ri]
\eea
Then the computation goes as follows (we denote for brevity $R:= \Pi_{\mathcal A}(\1_{\mathcal A} - K_{\mathcal A\mathcal A})^{-1}\Pi_{\mathcal A}$
\bea
\frac 1{ \det (\1_A- K_{\mathcal A \mathcal A}) }
 \det \le[ \begin{array}{c|ccc}
\1_{\mathcal A} - K_{\mathcal A \mathcal A} & -K_{\mathcal AE_1} &\dots & -K_{\mathcal AE_k}\nonumber\\
\hline
-\l_1K_{E_1\mathcal A} & \1_{E_1} -\l_1 K_{E_1E_1} & \dots &-\l_1 K_{E_1E_k}\nonumber\\
\vdots &&&\\
-\l_k K_{E_k \mathcal A} & -\l_kK_{E_k E_1} &\dots & \1_{E_k} - \l_k K_{E_kE_k}
\end{array}\ri] = \nonumber\\
 \det \le[ \begin{array}{c|ccc}
\1_{\mathcal A}  &  -RK_{\mathcal AE_1} &\dots & -R K_{\mathcal AE_k}\nonumber\\
\hline
-\l_1K_{E_1\mathcal A} & \1_{E_1} -\l_1 K_{E_1E_1} & \dots &-\l_1 K_{E_1E_k}\nonumber\\
\vdots &&&\\
-\l_k K_{E_k\mathcal A} & -\l_kK_{E_k E_1} &\dots & \1_{E_k} - \l_k K_{E_kE_k}
\end{array}\ri]=\nonumber\\
 \det
  \le[ \begin{array}{c|ccc}
\1_{\mathcal A}  & 0 &\dots & 0\\
\hline
\l_1K_{E_1\mathcal A} & \1_{E_1}  & \dots &0\\
\vdots &&&\\
\l_k K_{E_k\mathcal A} &0 &\dots & \1_{E_k} 
\end{array}\ri]
 \le[ \begin{array}{c|ccc}
\1_{\mathcal A}  & - RK_{\mathcal AE_1} &\dots &- R K_{\mathcal AE_k}\\
\hline
-\l_1K_{E_1\mathcal A} & \1_{E_1} -\l_1 K_{E_1E_1} & \dots &-\l_1 K_{E_1E_k}\\
\vdots &&&\\
-\l_k K_{E_k\mathcal A} & -\l_kK_{E_k E_1} &\dots & \1_{E_k} - \l_k K_{E_kE_k}
\end{array}\ri]=\nonumber\\
=
 \det
 \le[ \begin{array}{c|ccc}
\1_{\mathcal A}  & - RK_{\mathcal AE_1} &\dots & -R K_{\mathcal AE_k}\\
\hline
0 & \1_{E_1}  -\l_1 \le(K_{E_1E_1} + K_{E_1A}RK_{\mathcal AE_1}\ri)& \dots &-\l_1 \le(K_{E_1E_k}+ K_{E_1A}RK_{\mathcal AE_k}\ri)\\
\vdots &&&\\
0 &   -\l_1 \le(K_{E_kE_1} + K_{E_k\mathcal A}RK_{\mathcal AE_1}\ri)& \dots &\1_{E_k} -\l_1 \le(K_{E_kE_k}+K_{E_kA}RK_{\mathcal AE_k}\ri)\\
\end{array}\ri]=\nonumber\\
 \det
 \le[ \begin{array}{ccc}
 \1_{E_1}  -\l_1 \le(K_{E_1E_1} + K_{E_1\mathcal A}RK_{\mathcal AE_1}\ri)& \dots &-\l_1 \le(K_{E_1E_k}+ K_{E_1\mathcal A}RK_{\mathcal AE_k}\ri)\\
\vdots &&\\
   -\l_1 \le(K_{E_kE_1} +K_{E_k\mathcal A}RK_{\mathcal AE_1}\ri)& \dots &\1_{E_k} -\l_1 \le(K_{E_kE_k}+ K_{E_kA}RK_{\mathcal AE_k}\ri)\\
\end{array}\ri]\nonumber
\eea
The first determinant, $\det (\1_{\mathcal A}- K_{\mathcal A \mathcal A}) = F_{\mathcal A}(0)$ is the probability of finding no points in $\mathcal A$ and 
the last determinant is precisely the generating function of occupation numbers for the restricted process $K_{\mathcal A}$, which completes the proof.
\QED

\bibliographystyle{plain}
\bibliography{/Users/mattiacafasso/Documents/BibDeskLibrary.bib}
 \end{document}